\documentclass{article}

\usepackage{mathtools}
\usepackage{booktabs} 
\usepackage{etex}

\usepackage{authblk}
\usepackage{comment}
\usepackage{tikz}
\usetikzlibrary{matrix,arrows,decorations.pathmorphing,patterns,positioning}
\usepackage{amssymb}
\usepackage{ifthen}
\usepackage{color}
\usepackage{colortbl}
\usepackage{rotating}
\usepackage{multirow}
\usepackage{multicol}
\usepackage{url}
\usepackage{arydshln}

\usepackage{pgfplots}

\usetikzlibrary{external}
\usepackage{wrapfig}
\usepackage{subcaption}
\usepackage{array}
\usepackage{framed}

\usepackage{balance}

\usepackage{epsfig}
\usepackage{amsmath}
\usepackage{amsfonts}
\usepackage{amssymb}

\usepackage{algorithmic}

\usepackage{algorithm}

\usepackage[normalem]{ulem}

\usepackage{ifthen}
\usepackage{boxedminipage}
\usepackage{fancyvrb}

\usepackage{soul}
\usepackage{graphicx}
\usepackage{graphics}
\usepackage{moreverb}
\usepackage{rotating}
\usepackage{enumerate}

\usepackage{enumitem}

\usepackage{tabularx,booktabs}

\usetikzlibrary{arrows}

\usepackage{xspace}

\usepackage{stmaryrd}

\usepackage[absolute]{textpos}
\TPMargin{5pt}

\usepackage{listings}
\usepackage{xcolor}
\usepackage{fancyvrb}
\fvset{tabsize=2}
\definecolor{mygreen}{rgb}{0,0.6,0}
\usepackage{afterpage}

\usepackage[mathscr]{eucal}

\usepackage{bm}

\pgfplotscreateplotcyclelist{mycolorlist}{%
blue,every mark/.append style={fill=blue!80!black},mark=*\\%
red,every mark/.append style={fill=red!80!black},mark=square*\\%
brown!60!black,every mark/.append style={fill=brown!80!black},mark=otimes*\\%
black,mark=star\\%
blue,every mark/.append style={fill=blue!80!black},mark=diamond*\\%
red,densely dashed,every mark/.append style={solid,fill=red!80!black},mark=*\\%
brown!60!black,densely dashed,every mark/.append style={
solid,fill=brown!80!black},mark=square*\\%
black,densely dashed,every mark/.append style={solid,fill=gray},mark=otimes*\\%
blue,densely dashed,mark=star,every mark/.append style=solid\\%
red,densely dashed,every mark/.append style={solid,fill=red!80!black},mark=diamond*\\%
}

\definecolor{colorbrewer1}{RGB}{228,26,28}
\definecolor{colorbrewer2}{RGB}{55,126,184}
\definecolor{colorbrewer3}{RGB}{77,175,74}
\definecolor{colorbrewer4}{RGB}{152,78,163}
\definecolor{colorbrewer5}{RGB}{255,127,0}
\definecolor{colorbrewer7}{RGB}{166,86,40}
\definecolor{colorbrewer8}{RGB}{247,129,191}
\definecolor{colorbrewer9}{RGB}{153,153,153}

\definecolor{darkgreen}{rgb}{0,0.5,0}

\definecolor{darkred}{rgb}{0.44,0,0}
\definecolor{darkgreen}{rgb}{0,0.44,0}
\definecolor{darkblue}{rgb}{0,0,0.44}
\definecolor{enrique}{rgb}{0,0,0}

\pgfplotscreateplotcyclelist{jianyu2color}{
	colorbrewer1, mark=*,\\
	colorbrewer2, mark=square*,\\
	colorbrewer3, mark=triangle*,\\
	colorbrewer4, mark=x,\\
	colorbrewer5, mark=*,\\
	colorbrewer7, mark=triangle*,\\
	colorbrewer8, mark=+,\\
}

\pgfplotscreateplotcyclelist{jianyu0color}{%
thick,colorbrewer1,every mark/.append style={fill=colorbrewer1},mark=triangle*,only marks\\%
thick,orange,every mark/.append style={fill=orange},mark=square*,only marks\\%
thick, lime!80!black,every mark/.append style={fill=lime!80!black},mark=pentagon*,only marks\\%
thick,blue,every mark/.append style={fill=blue!80!black},mark=triangle*,only marks\\%
thick,cyan,every mark/.append style={fill=cyan},mark=square*,only marks\\%
thick,darkgreen,mark=pentagon*,only marks\\%
thick,yellow!60!black,every mark/.append style={solid,fill=yellow!80!black},mark=square*, only marks\\%
thick,purple,every mark/.append style={solid,fill=gray},mark=otimes*, only marks\\%
thick,blue,densely dashed,mark=star,every mark/.append style=solid, only marks\\%
thick,red,densely dashed,every mark/.append style={solid,fill=red!80!black},mark=diamond*, only marks\\%
thick,colorbrewer4,mark=x\\%
thick,blue,every mark/.append style={fill=blue!80!black},mark=*\\%
}

\pgfplotscreateplotcyclelist{jianyucolor}{%
thin,black,mark=o,only marks\\%
thick,colorbrewer1,every mark/.append style={fill=colorbrewer1},mark=triangle*,only marks\\%
thick,orange,every mark/.append style={fill=orange},mark=square*,only marks\\%
thick, lime!80!black,every mark/.append style={fill=lime!80!black},mark=pentagon*,only marks\\%
thick,blue,every mark/.append style={fill=blue!80!black},mark=triangle*,only marks\\%
thick,cyan,every mark/.append style={fill=cyan},mark=square*,only marks\\%
thick,darkgreen,mark=pentagon*,only marks\\%
thick,yellow!60!black,every mark/.append style={solid,fill=yellow!80!black},mark=square*, only marks\\%
thick,purple,every mark/.append style={solid,fill=gray},mark=otimes*, only marks\\%
thick,blue,densely dashed,mark=star,every mark/.append style=solid, only marks\\%
thick,red,densely dashed,every mark/.append style={solid,fill=red!80!black},mark=diamond*, only marks\\%
thick,colorbrewer4,mark=x\\%
thick,blue,every mark/.append style={fill=blue!80!black},mark=*\\%
red,every mark/.append style={fill=red!80!black},mark=square*\\%
brown!60!black,every mark/.append style={fill=brown!80!black},mark=diamond*\\%
blue,every mark/.append style={fill=blue!80!black},mark=diamond*\\%
thick,red,densely dashed,every mark/.append style={solid,fill=red!80!black},mark=*, only marks\\%
red,densely dashed,every mark/.append style={solid,fill=red!80!black},mark=*\\%
brown!60!black,densely dashed,every mark/.append style={
solid,fill=brown!80!black},mark=square*\\%
darkred,densely dashed,every mark/.append style={solid,fill=gray},mark=otimes*\\%
blue,densely dashed,mark=oplus,every mark/.append style=solid\\%
colorbrewer5,densely dashed,every mark/.append style={solid,fill=colorbrewer5},mark=diamond*\\%
darkgreen,densely dashed,mark=+,every mark/.append style=solid\\%
thick,colorbrewer7,mark=triangle*\\%
thick,black,densely dashed\\%
thick,black\\%
}

\pgfplotscreateplotcyclelist{jianyucolorbar}{%
fill=black\\%
fill=colorbrewer1\\%
fill=orange\\%
fill=lime!80!black\\%
fill=blue\\%
fill=cyan\\%
fill=darkgreen\\%
fill=yellow!60!black\\%
thick,purple,every mark/.append style={solid,fill=gray}\\%
thick,blue,densely dashed,mark=star,every mark/.append style=solid\\%
thick,red,densely dashed,every mark/.append style={solid,fill=red!80!black},mark=diamond*, only marks\\%
thick,colorbrewer4,mark=x\\%
thick,blue,every mark/.append style={fill=blue!80!black},mark=*\\%
red,every mark/.append style={fill=red!80!black},mark=square*\\%
brown!60!black,every mark/.append style={fill=brown!80!black},mark=diamond*\\%
blue,every mark/.append style={fill=blue!80!black},mark=diamond*\\%
thick,red,densely dashed,every mark/.append style={solid,fill=red!80!black},mark=*, only marks\\%
red,densely dashed,every mark/.append style={solid,fill=red!80!black},mark=*\\%
brown!60!black,densely dashed,every mark/.append style={
solid,fill=brown!80!black},mark=square*\\%
darkred,densely dashed,every mark/.append style={solid,fill=gray},mark=otimes*\\%
blue,densely dashed,mark=oplus,every mark/.append style=solid\\%
colorbrewer5,densely dashed,every mark/.append style={solid,fill=colorbrewer5},mark=diamond*\\%
darkgreen,densely dashed,mark=+,every mark/.append style=solid\\%
thick,colorbrewer7,mark=triangle*\\%
thick,black,densely dashed\\%
thick,black\\%
}

\pgfplotscreateplotcyclelist{jianyucolorlineratio}{%
thick,black,mark=*\\%
thick,colorbrewer1,every mark/.append style={fill=colorbrewer1},mark=triangle*\\%
thick,orange,every mark/.append style={fill=orange},mark=square*\\%
thick, lime!80!black,every mark/.append style={fill=lime!80!black},mark=pentagon*\\%
thick,blue,every mark/.append style={fill=blue!80!black},mark=triangle*\\%
thick,cyan,every mark/.append style={fill=cyan},mark=square*\\%
thick,darkgreen,mark=pentagon*\\%
thick,yellow!60!black,every mark/.append style={solid,fill=yellow!80!black}\\%
thick,purple,every mark/.append style={solid,fill=gray}\\%
thick,blue,densely dashed,mark=star,every mark/.append style=solid\\%
thick,red,densely dashed,every mark/.append style={solid,fill=red!80!black},mark=diamond*, only marks\\%
thick,colorbrewer4,mark=x\\%
thick,blue,every mark/.append style={fill=blue!80!black},mark=*\\%
red,every mark/.append style={fill=red!80!black},mark=square*\\%
brown!60!black,every mark/.append style={fill=brown!80!black},mark=diamond*\\%
blue,every mark/.append style={fill=blue!80!black},mark=diamond*\\%
thick,red,densely dashed,every mark/.append style={solid,fill=red!80!black},mark=*, only marks\\%
red,densely dashed,every mark/.append style={solid,fill=red!80!black},mark=*\\%
brown!60!black,densely dashed,every mark/.append style={
solid,fill=brown!80!black},mark=square*\\%
darkred,densely dashed,every mark/.append style={solid,fill=gray},mark=otimes*\\%
blue,densely dashed,mark=oplus,every mark/.append style=solid\\%
colorbrewer5,densely dashed,every mark/.append style={solid,fill=colorbrewer5},mark=diamond*\\%
darkgreen,densely dashed,mark=+,every mark/.append style=solid\\%
thick,colorbrewer7,mark=triangle*\\%
thick,black,densely dashed\\%
thick,black\\%
}

\pgfplotscreateplotcyclelist{jianyucolorline}{%
thin,black\\%
thick,colorbrewer1,every mark/.append style={fill=colorbrewer1}\\%
thick,orange,every mark/.append style={fill=orange}\\%
thick, lime!80!black,every mark/.append style={fill=lime!80!black}\\%
thick,blue,every mark/.append style={fill=blue!80!black}\\%
thick,cyan,every mark/.append style={fill=cyan}\\%
thick,darkgreen\\%
thick,yellow!60!black,every mark/.append style={solid,fill=yellow!80!black}\\%
thick,purple,every mark/.append style={solid,fill=gray}\\%
thick,blue,densely dashed,mark=star,every mark/.append style=solid\\%
thick,red,densely dashed,every mark/.append style={solid,fill=red!80!black},mark=diamond*, only marks\\%
thick,colorbrewer4,mark=x\\%
thick,blue,every mark/.append style={fill=blue!80!black},mark=*\\%
red,every mark/.append style={fill=red!80!black},mark=square*\\%
brown!60!black,every mark/.append style={fill=brown!80!black},mark=diamond*\\%
blue,every mark/.append style={fill=blue!80!black},mark=diamond*\\%
thick,red,densely dashed,every mark/.append style={solid,fill=red!80!black},mark=*, only marks\\%
red,densely dashed,every mark/.append style={solid,fill=red!80!black},mark=*\\%
brown!60!black,densely dashed,every mark/.append style={
solid,fill=brown!80!black},mark=square*\\%
darkred,densely dashed,every mark/.append style={solid,fill=gray},mark=otimes*\\%
blue,densely dashed,mark=oplus,every mark/.append style=solid\\%
colorbrewer5,densely dashed,every mark/.append style={solid,fill=colorbrewer5},mark=diamond*\\%
darkgreen,densely dashed,mark=+,every mark/.append style=solid\\%
thick,colorbrewer7,mark=triangle*\\%
thick,black,densely dashed\\%
thick,black\\%
}

\pgfplotscreateplotcyclelist{jianyucolordistline}{%
thin,black,mark=*\\%
thin,colorbrewer1,every mark/.append style={fill=colorbrewer1},mark=triangle*\\%
thin,orange,every mark/.append style={fill=orange},mark=square*\\%
thin,lime!80!black,every mark/.append style={fill=lime!80!black},mark=pentagon*\\%
thin,blue,every mark/.append style={fill=blue!80!black},mark=triangle*\\%
thin,cyan,every mark/.append style={fill=cyan},mark=square*\\%
thin,darkgreen,mark=pentagon*\\%
densely dashed,black,mark=x\\%
thick,yellow!60!black,every mark/.append style={solid,fill=yellow!80!black}\\%
thick,purple,every mark/.append style={solid,fill=gray}\\%
thick,blue,densely dashed,mark=star,every mark/.append style=solid\\%
thick,red,densely dashed,every mark/.append style={solid,fill=red!80!black},mark=diamond*, only marks\\%
thick,colorbrewer4,mark=x\\%
thick,blue,every mark/.append style={fill=blue!80!black},mark=*\\%
red,every mark/.append style={fill=red!80!black},mark=square*\\%
brown!60!black,every mark/.append style={fill=brown!80!black},mark=diamond*\\%
blue,every mark/.append style={fill=blue!80!black},mark=diamond*\\%
thick,red,densely dashed,every mark/.append style={solid,fill=red!80!black},mark=*, only marks\\%
red,densely dashed,every mark/.append style={solid,fill=red!80!black},mark=*\\%
brown!60!black,densely dashed,every mark/.append style={
solid,fill=brown!80!black},mark=square*\\%
darkred,densely dashed,every mark/.append style={solid,fill=gray},mark=otimes*\\%
blue,densely dashed,mark=oplus,every mark/.append style=solid\\%
colorbrewer5,densely dashed,every mark/.append style={solid,fill=colorbrewer5},mark=diamond*\\%
darkgreen,densely dashed,mark=+,every mark/.append style=solid\\%
thick,colorbrewer7,mark=triangle*\\%
thick,black,densely dashed\\%
thick,black\\%
}

\pgfplotscreateplotcyclelist{jianyucolorboth}{%
thin,black,mark=o,only marks\\%
thick,colorbrewer1,every mark/.append style={fill=colorbrewer1},mark=triangle*,only marks\\%
thick,orange,every mark/.append style={fill=orange},mark=square*,only marks\\%
thick, lime!80!black,every mark/.append style={fill=lime!80!black},mark=pentagon*,only marks\\%
thick,blue,every mark/.append style={fill=blue!80!black},mark=triangle*,only marks\\%
thick,cyan,every mark/.append style={fill=cyan},mark=square*,only marks\\%
thick,darkgreen,mark=pentagon*,only marks\\
thin,black\\%
thick,colorbrewer1,every mark/.append style={fill=colorbrewer1}\\%
thick,orange,every mark/.append style={fill=orange}\\%
thick, lime!80!black,every mark/.append style={fill=lime!80!black}\\%
thick,blue,every mark/.append style={fill=blue!80!black}\\%
thick,cyan,every mark/.append style={fill=cyan}\\%
thick,darkgreen\\
thick,yellow!60!black,every mark/.append style={solid,fill=yellow!80!black}\\%
thick,purple,every mark/.append style={solid,fill=gray}\\%
thick,blue,densely dashed,mark=star,every mark/.append style=solid\\%
thick,red,densely dashed,every mark/.append style={solid,fill=red!80!black},mark=diamond*, only marks\\%
thick,colorbrewer4,mark=x\\%
thick,blue,every mark/.append style={fill=blue!80!black},mark=*\\%
red,every mark/.append style={fill=red!80!black},mark=square*\\%
brown!60!black,every mark/.append style={fill=brown!80!black},mark=diamond*\\%
blue,every mark/.append style={fill=blue!80!black},mark=diamond*\\%
thick,red,densely dashed,every mark/.append style={solid,fill=red!80!black},mark=*, only marks\\%
red,densely dashed,every mark/.append style={solid,fill=red!80!black},mark=*\\%
brown!60!black,densely dashed,every mark/.append style={
solid,fill=brown!80!black},mark=square*\\%
darkred,densely dashed,every mark/.append style={solid,fill=gray},mark=otimes*\\%
blue,densely dashed,mark=oplus,every mark/.append style=solid\\%
colorbrewer5,densely dashed,every mark/.append style={solid,fill=colorbrewer5},mark=diamond*\\%
darkgreen,densely dashed,mark=+,every mark/.append style=solid\\%
thick,colorbrewer7,mark=triangle*\\%
thick,black,densely dashed\\%
thick,black\\%
}

\pgfplotscreateplotcyclelist{jianyu3color}{%
black,densely dashed\\%
black\\%
colorbrewer1,every mark/.append style={fill=colorbrewer1},mark=triangle*\\%
orange,every mark/.append style={fill=orange},mark=square*\\%
cyan,every mark/.append style={fill=cyan},mark=otimes*\\%
red!70!white,mark=star\\%
brown!60!black,every mark/.append style={fill=brown!80!black},mark=otimes*\\%
purple,every mark/.append style={solid,fill=gray},mark=otimes*\\%
very thick, blue,every mark/.append style={fill=lime},mark=diamond*\\%
very thick, darkgreen,mark=+,every mark/.append style=solid\\%
}

\pgfplotscreateplotcyclelist{jianyu4color}{%
black,densely dashed,mark=triangle*\\%
black,mark=triangle*\\%
colorbrewer1,densely dashed,mark=triangle*\\%
thick,colorbrewer1,mark=triangle*\\%
black,densely dashed,mark=square*\\%
black,mark=square*\\%
blue,densely dashed,mark=square*\\%
thick,blue,mark=square*\\%
black,densely dashed,mark=*\\%
black,mark=*\\%
darkgreen,densely dashed,mark=*\\%
thick,darkgreen,mark=*\\%
orange,mark=square\\%
cyan,mark=otimes\\%
purple,mark=star\\%
}

\pgfplotscreateplotcyclelist{jianyu5color}{%
black,densely dashed\\%
black\\%
darkgreen,densely dashed,mark=*\\%
thick,darkgreen,mark=*\\%
blue,densely dashed,mark=square*\\%
thick,blue,mark=square*\\%
colorbrewer1,densely dashed,mark=triangle*\\%
thick,colorbrewer1,mark=triangle*\\%
}

\pgfplotscreateplotcyclelist{jianyu6color}{%
black,densely dashed\\%
black\\%
very thick,densely dashed,colorbrewer1,every mark/.append style={fill=colorbrewer1},mark=triangle*\\%
very thick,orange,every mark/.append style={fill=orange},mark=square*\\%
cyan,every mark/.append style={fill=cyan},mark=otimes*\\%
red!70!white,mark=star\\%
brown!60!black,every mark/.append style={fill=brown!80!black},mark=otimes*\\%
purple,every mark/.append style={solid,fill=gray},mark=otimes*\\%
blue,every mark/.append style={fill=lime},mark=diamond*\\%
darkgreen,mark=+,every mark/.append style=solid\\%
}

\newcommand{\jianyuFromTo}[2]{#2}

\newcommand{\ABCstrassen}{\textbf{ABC \mbox{Strassen}}}
\newcommand{\ABXstrassen}{\textbf{AB \mbox{Strassen}}}
\newcommand{\XXXstrassen}{\textbf{Naive \mbox{Strassen}}}

\newcommand{\figref}[1]{Figure~\ref{#1}}

\newcommand{\secref}[1]{$\S$\ref{#1}}

\newcommand{\strassen}{\mbox{\sc Strassen}}
\newcommand{\ttdt}{\mbox{\sc ttdt}}

\newcommand{\GOTO}{{\sc GotoBLAS}}
\newcommand{\BLIS}{{\sc BLIS}}

\newcommand{\gemm}{{\sc gemm}\xspace}

\newcommand{\dgemm}{{\sc dgemm}}

\newcommand{\tblis}{{\sc tblis}\xspace}

\newcommand{\ttt}{{\sc ttt}\xspace}
\newcommand{\ctf}{{\sc ctf}\xspace}

\newcommand{\NImCap}{{\widetilde {N_{I_m}}}}
\newcommand{\NJnCap}{{\widetilde {N_{J_n}}}}
\newcommand{\NPkCap}{{\widetilde {N_{P_k}}}}

\newcommand{\NIm}{{ {N_{I_m}}}}
\newcommand{\NJn}{{ {N_{J_n}}}}
\newcommand{\NPk}{{ {N_{P_k}}}}

\newcommand{\coeffa}{ W_{a} }
\newcommand{\coeffm}{ W_{m} }

\newcommand{\fromto}[2]{{\color{black} #2}}

\newcolumntype{I}{!{\vrule width 1.5pt}}
\newlength\savedwidth
\newcommand\whline{\noalign{\global\savedwidth\arrayrulewidth
                            \global\arrayrulewidth 1.5pt}%
           \hline
           \noalign{\global\arrayrulewidth\savedwidth}}

\newcommand{\DivisorML}{ 2^L }
\newcommand{\DivisorNL}{ 2^L }
\newcommand{\DivisorKL}{ 2^L }

\newcommand{\SQUARE}{$ \NIm{}\!\!\approx\!\! \NJn{} \!\!\approx\!\! \NPk{} $}
\newcommand{\RANKK}{$ \NIm{} \!\!\approx\!\! \NJn{} \!\!\approx\!\! 16000 $, $ \NPk{} $ varies}
\newcommand{\FIXK}{$ \NPk{} \!\!\approx\!\! 1024 $, $ \NIm{} \!\! \approx\!\! \NJn{} $ vary}

\newcommand{\mycircle}[1]{{\textcircled{\raisebox{-0.9pt}{#1}}}}

\newcommand{\T}[1]{\bm{\mathcal{\MakeUppercase{#1}}}}

\newcommand*{\affaddr}[1]{#1} 
\newcommand*{\affmark}[1][*]{\textsuperscript{#1}}
\newcommand*{\email}[1]{\texttt{#1}}

\date{
April 3, 2017
}

\begin{document}

\title{
\Large Strassen's Algorithm for Tensor Contraction
\\[0.2in]
\large FLAME Working Note \#84
}

\author{
Jianyu Huang\affmark[*]\affmark[\dag],
Devin A. Matthews\affmark[\dag],
Robert A. van de Geijn\affmark[*]\affmark[\dag]\\
\affaddr{\affmark[*]Department of Computer Science}\\
\affaddr{\affmark[\dag]Institute for Computational Engineering and Sciences}\\
\affaddr{The University of Texas at Austin, Austin, TX 78712}\\
\email{{\tt \{jianyu@cs.}, {\tt dmatthews@}, {\tt rvdg@cs.\}utexas.edu}}
}

\newcommand{\NoShow}[1]{}


\maketitle

\begin{abstract}

Tensor contraction (TC) is an important computational kernel widely used in numerous applications.
It is a multi-dimensional generalization of matrix multiplication (GEMM). While Strassen's algorithm for GEMM is well studied in theory and practice, extending it to accelerate TC has not been previously pursued.
Thus, we believe this to be the first paper to demonstrate how one can in practice speed up tensor contraction with Strassen's algorithm.
By adopting a Block-Scatter-Matrix format, a novel matrix-centric tensor layout, we can conceptually view TC as GEMM for a general stride storage, with an implicit tensor-to-matrix transformation.
This insight enables us to tailor a recent state-of-the-art implementation of Strassen's algorithm to TC,
avoiding explicit transpositions (permutations) and extra workspace, and reducing the overhead of memory movement that is incurred.
Performance benefits are demonstrated with a performance model as well as in practice on modern single core, multicore, and distributed memory parallel architectures, achieving up to $ 1.3 \times $ speedup.
The resulting implementations can serve as a drop-in replacement for various applications with significant speedup.

\end{abstract}

\section{Introduction}

\noindent{\bf Standing on the shoulders of giants.}
This paper builds upon a number of recent developments:
The GotoBLAS algorithm for matrix multiplication (GEMM)~\cite{Goto:2008:AHP}
that underlies the currently fastest implementations of GEMM for CPUs; The
refactoring of the GotoBLAS algorithm as part of the BLAS-like Library
Instantiation Software (BLIS)~\cite{BLIS1,BLIS2}, which exposes primitives for
implementing BLAS-like operations; The systematic parallelization of the loops
that BLIS exposes so that  high-performance can be flexibly attained on multicore
and many-core architectures~\cite{BLIS3}; The casting of tensor contraction (TC)
in terms of the BLIS primitives~\cite{TC:Devin,Paul16} without requiring the
transposition (permutation) used by traditional implementations;  The practical
high-performance implementation of the classical Strassen's algorithm
(\strassen)~\cite{Strassen:SC16} in terms of variants of the BLIS primitives; and
the extension of this implementation~\cite{FMM:IPDPS17} to a family of
Strassen-like algorithms (Fast Matrix Multiplication algorithms)~\cite{Benson15}.
Together, these results facilitate what we believe to be the first extension of
Strassen's algorithm to TC.

\noindent{\bf Contributions.}
This paper describes how to extend Strassen's algorithm to TC without the explicit
transposition of data that inherently incurs significant memory movement and
workspace overhead;  It provides a performance model for the cost of the
resulting family of algorithms; It details the practical implementation of these
algorithms, including how to exploit variants of the primitives that underlie
BLIS and a data layout to memory for the tensors;  It demonstrates practical
speedup on modern single core and multicore CPUs;  It illustrates how the local
Strassen's  TC algorithm improves performance of a simple distributed memory
tensor contraction.
Together, these results unlock a new frontier for the research and application
of Strassen's algorithm.


\label{sec:related}
\noindent{\bf Related work.}
To the best of our knowledge, this work represents the first implementation of Strassen's algorithm for tensor contraction.
In the context of \strassen\ for matrices, there have been a variety of practical implementations \cite{StrassenDouglas,Huss-Lederman:1996:ISA:369028.369096,D'alberto:2011:EPM:2049662.2049664,Benson15}, including the closely
related implementation of \strassen\ using the BLIS framework \cite{Strassen:SC16}
which this paper is based on.

For tensor contraction, recent work on high-performance tensor contraction \cite{TC:Devin,Paul16} serves as the
motivation and basis for our present work, while other research has focused on algorithms using tensor
slicing \cite{slice1,slice2,slice3,slice_gpu1} or on improving the efficiency of the
so-called \ttdt\ algorithm for tensor contraction \cite{hartono_performance_2009,hanrath_efficient_2010,lyakh_efficient_2015,ttc}, where input tensors $\T{A}$ and $\T{B}$
are \underline{T}ransposed (permuted) and then used in a standard \textsc{\underline{d}gemm} algorithm, with the output then being \underline{T}ransposed
and accumulated onto the tensor $\T{C}$. \ttdt\ could be used to construct a \strassen\ algorithm for TC by
transposing subtensors into submatrices and vice versa and using a matrix implementation of \strassen\ instead
of \dgemm{}. However, we will show that this algorithm is essentially the same as our \XXXstrassen\ algorithm,
which is often less efficient than the other algorithms that we have implemented.

The {\sc gett} algorithm \cite{Paul16} is a high-performance tensor contraction implementation similar in
many ways to the BLIS-based implementation in \cite{TC:Devin}. As in
\cite{TC:Devin}, which our present work is based on, 
formation of linear combinations of input subtensors of $\T{A}$ and $\T{B}$ and output to multiple subtensors of $\T{C}$ could
be fused with the internal tensor transposition and micro-kernel steps of {\sc gett}. However, the implementation
would be restricted to regular subtensors rather than more general submatrices, which could have possible
negative performance implications.


\section{Background}

We briefly review how high-performance GEMM is implemented, before
discussing the practical implementations of high-performance \strassen\ for GEMM.

\subsection{High-performance GEMM}
\label{sec:gemm}

\begin{figure*}[tb!]
~
\vspace{-0.60in}
\begin{center}
\includegraphics[width=1.03\textwidth]{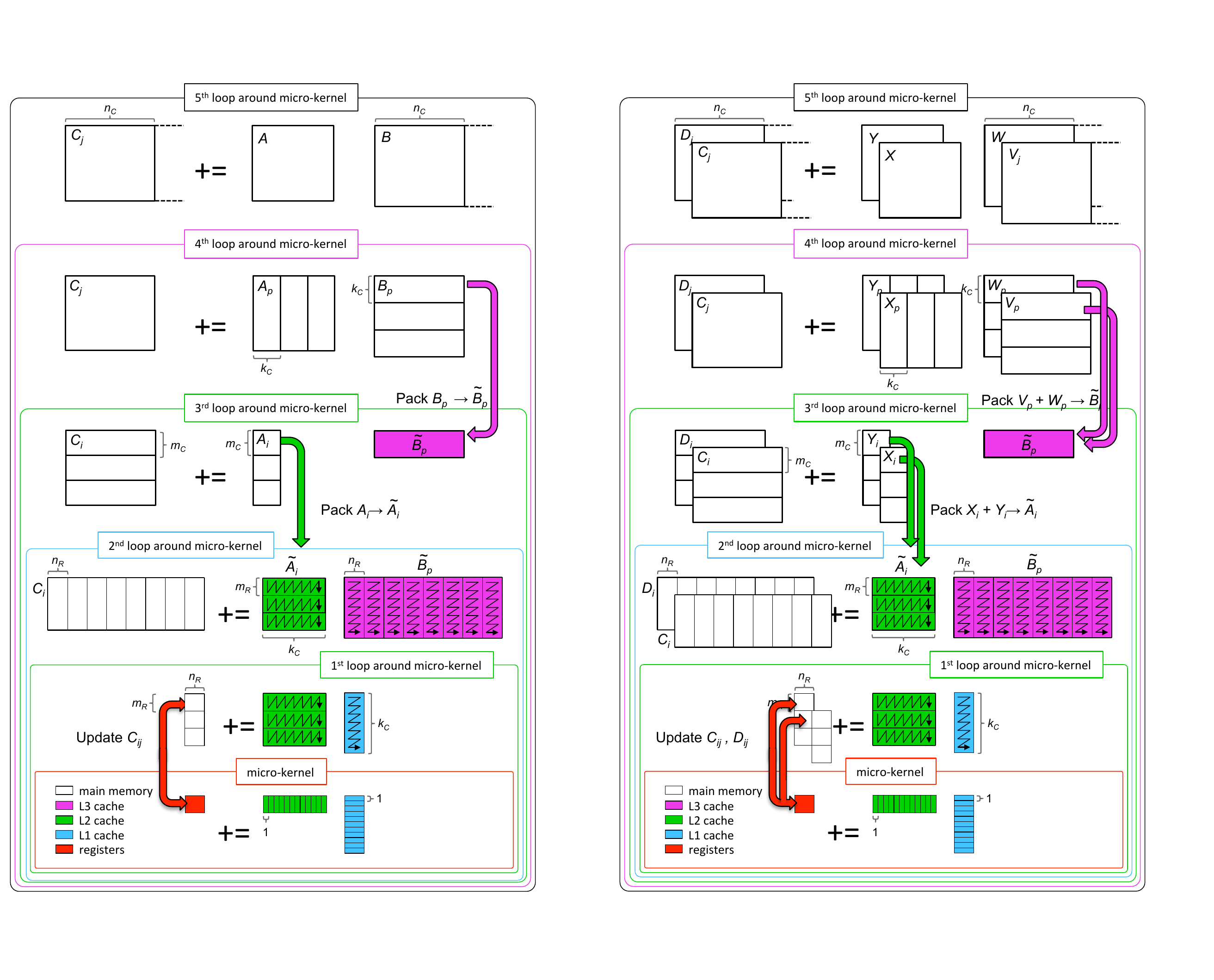}
\end{center}
\vspace{-0.55in}
\caption{
	Figure from~\cite{Strassen:SC16} (used with permission from authors).
	Left: illustration of the BLIS implementation of the \GOTO{} \gemm\  algorithm.  All computation is cast in terms of a highly optimized micro-kernel.
    Right: modification that implements the representative computation $ M = ( X + Y)( V+W); C +\!\!= M; D +\!\!= M $ of each row of computations in (\ref{eqn:allops}).
$ X $, $ Y $ are submatrices of $ A $; $ V $, $ W $ are submatrices of $ B $; $ C $, $ D $ are submatrices of the original matrix $ C $; $ M $ is the intermediate matrix product. Note that the packing buffers $ \widetilde A_i $ and $ \widetilde B_p $ stay in cache.
}
\label{fig:side_by_side}
\end{figure*}

Let  $ A $, $ B $, and $ C $ be matrices of sizes $ N_i \times N_p $, $ N_p \times N_j $, and
$N_i \times N_j $, respectively. A general matrix-matrix multiplication (\gemm{}) in the BLAS
interface \cite{BLAS3} is expressed as $ C := \alpha A B + \beta C $. Written element-wise,
$C_{i,j} = \alpha \sum_{p=0}^{N_p-1} A_{i,p} \cdot B_{p,j} + \beta C_{i,j}$,
where $\cdot$ denotes scalar multiplication, and $ \alpha $ and $ \beta $ are scalars.
We focus on the special case $ \alpha = 1 $ and $ \beta = 1 $ henceforth for brevity.

A key insight underlying modern high-performance implementations of \gemm{} is to organize
the computations by partitioning the operands into blocks for temporal locality, and to \emph{pack} (copy)
such blocks into contiguous buffers that fit into various levels of memory for spatial locality.
\figref{fig:side_by_side}(left) illustrates the \GOTO\ algorithm as implemented in BLIS.
Cache blocking parameters $\{m_C, n_C, k_C\}$ determine the submatrix sizes of $ B_p $ ($ k_C \times n_C $) and $ A_i $ ($ m_C \times k_C$),
such that they fit in various caches (we use the standard \gemm{} dimensions $\{m, n, k\}$ in defining
blocking parameters for brevity and consistency with \cite{BLIS1}, but note that the meaning of $\{m, n, k\}$ alone is
changed in \secref{sec:tensor}).
During the computation, row panels $ B_p $ are contiguously packed into buffer $ \widetilde B_p $ to fit in the L3 cache.
Blocks $ A_i $ are similarly packed into buffer $ \widetilde A_i $ to fit in the L2 cache.
Register block sizes $\{m_R, n_R\}$ relate to submatrices in registers
that contribute to $ C $.
In the \emph{micro-kernel} (the inner most loop),
a small $ m_R \times n_R $ micro-tile of $ C $ is updated by pair of $m_R \times k_C $ and $ k_C \times n_R $ slivers
of $ \widetilde A_i $ and $ \widetilde B_p $.
The above parameters can be analytically chosen~\cite{BLIS4}.

\subsection{High-performance \strassen}
\label{sec:strassen}


If the three operands are partitioned into quadrants,
\begin{equation}
\label{eqn:strassenpart}
X =
\left( \begin{array}{c | c}
X_{0} & X_{1} \\ \hline
X_{2} & X_{3}
\end{array}\right)
\quad \mbox{for $ X \in \{ A, B, C\} $}
\end{equation}
then it can be checked that the operations
\begin{equation}
{
\begin{array}{l @{\hspace{1pt}} c @{\hspace{1pt}} l l r}
M_0 &=&  ( A_{0} + A_{3} ) ( B_{0} + B_{3} );
&
C_{0} +\!\!= M_0;  C_{3} +\!\!= M_0;  \\
M_1 &=&  ( A_{2} + A_{3} ) B_{0};
&
C_{2} +\!\!= M_1 ;  C_{3} -\!\!= M_1 ; \\
M_2 &=&  A_{0} ( B_{1} - B_{3} );
&
C_{1} +\!\!= M_2 ;  C_{3} +\!\!= M_2 ;
\\
M_3 &=&  A_{3}( B_{2} - B_{0} );
&
C_{0} +\!\!= M_3 ;  C_{2} +\!\!= M_3 ;
\\
M_4 &=&  ( A_{0} + A_{1}) B_{3};
&
C_{1} +\!\!=  M_4 ;   C_{0} -\!\!= M_4;
\\
M_5&=&  (A_{2} - A_{0} )( B_{0} + B_{1} );
&
C_{3} +\!\!= M_5;
\\
M_6&=&  (A_{1} - A_{3} )( B_{2} + B_{3} );
&
C_{0} +\!\!= M_6 ;
\end{array}
}
\label{eqn:allops}
\end{equation} %
compute $ C := A B + C $, with seven instead of eight (sub)matrix multiplications,
reducing the cost by a factor of $ 7 / 8 $ (ignoring a lower order number of extra additions).
If all matrices are square and of size $ N \times N $,
theoretically this single step of \strassen\ can be  applied recursively, resulting in the classical \strassen\ with a cost of $ O ( N^{2.801} ) $.

In practice, only a few levels of the recursion are leveraged because
the reduction in computations are quickly overwhelmed by the cost of extra additions and extra memory movements.
Additionally, \strassen\ is known to experience degradation in numerical stability especially when more than two levels of
recursion are incorporated \cite{HighamBook,DemmelStrassenStable2007,Ballard15}.


\figref{fig:side_by_side}(right)
illustrates the modifications done in \cite{Strassen:SC16} to make \strassen\ practical.
During the \emph{packing} process,
the additions of the submatrices $ A $ and $ B $ can be incorporated into the packing buffers $ \widetilde A_i $ and $ \widetilde B_p $,
avoiding extra memory movement and reducing workspace requirements.
In the \emph{micro-kernel},
once a submatrix that contributes to $ C $ is computed in machine registers, it can be
directly added to the appropriate parts of multiple submatrices of $ C $,
thus avoiding the need for temporary intermediate matrices $ M_i $, again avoiding extra memory movement.
As demonstrated in~\cite{Strassen:SC16}, this approach makes 
\strassen\ practical for smaller matrices
and matrices of special shape (importantly, for rank-k updates, where $ N_p $ is relatively small comparing to $ N_i $ and $ N_j $).
This research is pushed further~\cite{FMM:IPDPS17} by revealing that
\strassen\ performs relatively better than most other Strassen-like FMM algorithms with one or two levels of recursions,
when modeled as well as in practice.  For this reason, we do not extend those FMM algorithms to TC in this paper, although
it may be worthwhile in future work to pursue certain of these algorithms for highly non-square tensor contraction
shapes.

\subsection{High-performance Tensor Contraction}

\label{sec:tensor}

The definition and notation of tensors and tensor contraction are briefly reviewed
before describing the tensor layouts that enable high-performance tensor contraction.

\noindent
{\bf Tensor.} The concept of matrices is extended to multiple dimensions by defining a
general $d$-D tensor $ \T{T} \in \mathbb{R}^{N_{\T{T};0} \times
\dots \times N_{\T{T};d-1} } $ as a multidimensional array of scalar elements,
where the length of the $k$-th dimension is given by $N_{\T{T};k} \in \mathbb{N}$.
Individual elements are referenced by indexing $\T{T}$ by an ordered \emph{index bundle}
$ T_d = \{t_0, \ldots, t_{d-1} \} $, such that $\T{T}_{T_d} \in \mathbb{R} $ for all 
$ T_d \in N_{\T{T};0} \times \ldots \times N_{\T{T};d-1} = N_{t_0} \times \ldots
\times N_{t_{d-1}}$. 
In general we will denote the dimension of a tensor $\T{T}$ as $d_{\T{T}}$, the index
length $N_x \in \mathbb{N}$ as the length of the dimension that is indexed
by some symbol $x$, and the bundle length $ N_{T_{d}} \in \mathbb{N} $ as the total length of
a index bundle $ T_d $, i.e. $ N_{T_{d}} = \prod_{t \in T_d} N_t = N_{t_0} \cdot \ldots \cdot N_{t_{d-1}} $.

\noindent
{\bf Tensor Contraction.}
Let $\T{A}$, $\T{B}$, and $\T{C}$ be general tensors of any dimensionality
satisfying $d_{\T{A}} + d_{\T{B}} - d_{\T{C}} = 2 k,\; k \in \mathbb{N}$.
Then, let $I_m$, $J_n$, and $P_k$ be index bundles with $m = d_{\T{A}}-k$ and
$n = d_{\T{B}}-k$. Lastly, let the index reordering $\pi_{\T{A}}(a_0,\ldots,a_{d_{\T{A}}-1})
= \{a_{\pi_{\T{A}}(0)},\ldots,a_{\pi_{\T{A}}(d_{\T{A}}-1)}\}$ be defined by the bijective map
$\pi_{\T{A}} \colon \{0,\ldots,d_{\T{A}}-1\} \to \{0,\ldots,d_{\T{A}}-1\}$,
and similarly
for $\pi_{\T{B}}$ and $\pi_{\T{C}}$. The general definition of tensor contraction is then given by,
\begin{align*}
\T{C}_{\pi_{\T{C}}(I_m J_n)} \coloneqq & \,\alpha \sum\limits_{P_k \in N_{p_0}\times\ldots\times N_{p_{k-1}}} 
\T{A}_{\pi_{\T{A}}(I_m P_k)}  \cdot \T{B}_{\pi_{\T{B}}(P_k J_n)} + \beta  \T{C}_{\pi_{\T{C}}(I_m J_n)},
\end{align*}
for scalars $\alpha,\,\beta \in \mathbb{R}$. The indices in the bundles $I_n$ and $J_m$ are generally
called \emph{free}, \emph{external}, or \emph{uncontracted} indices, while the indices in the $P_k$ bundle are called \emph{bound}, \emph{internal}, or
\emph{contracted} indices. In the following we will assume that $\alpha=1$ and $\beta=1$, and suppress the explicit summation
over $P_k$.
The number of leading-order floating point operations required for tensor contraction is $ 2 \NIm{} \cdot \NJn{} \cdot \NPk{} $
= $2 \prod_{i \in I_m} N_i \cdot \prod_{j \in J_n} N_j \cdot \prod_{p \in P_k} N_p $.
If the length of each dimension is $O(N)$, the tensor contraction operation requires
$ O( N^{m+n+k} ) $ flops.

\begin{figure*}[tb!]
~
\begin{subfigure}[b]{1\textwidth}
\centering
\includegraphics[width=1.06\textwidth]{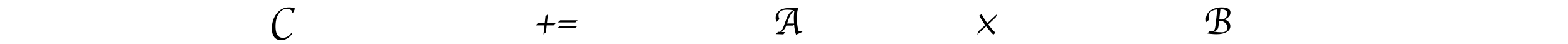}
\includegraphics[width=1.06\textwidth]{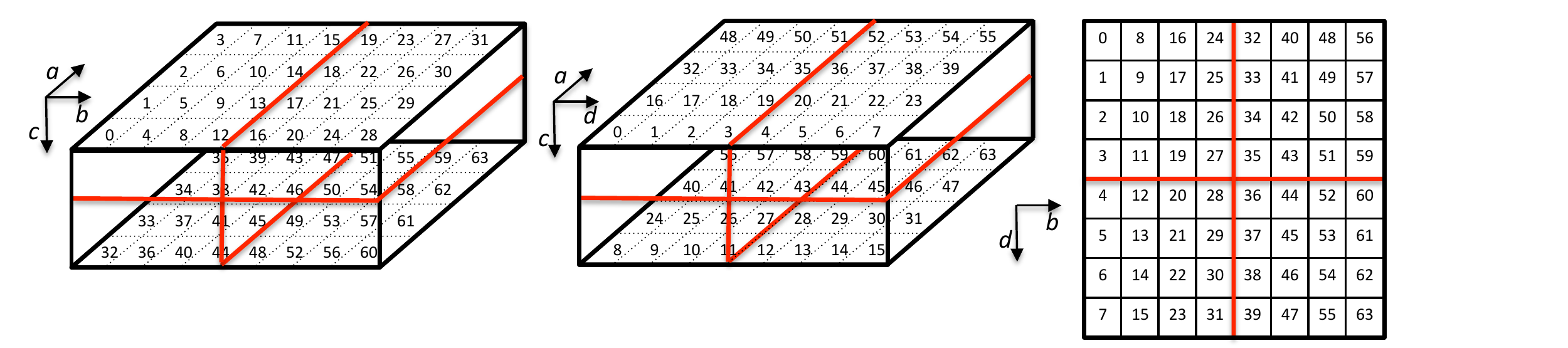}
\vspace{-0.2in}
\caption{Tensor contraction $ \T{C}_{a,b,c} +\!\!= \T{A}_{d,c,a} \cdot \T{B}_{d,b} $ with $N_a=4$, $N_b=N_d=8$, and $N_c=2$. The
relative location of each data element in memory is given assuming a generalized column-major layout.
}
\label{fig:tensor_contraction}
\end{subfigure}
\begin{subfigure}[b]{1\textwidth}
\centering
\includegraphics[width=1.06\textwidth]{figures/stra_tensor_6.pdf}
\includegraphics[width=1.06\textwidth]{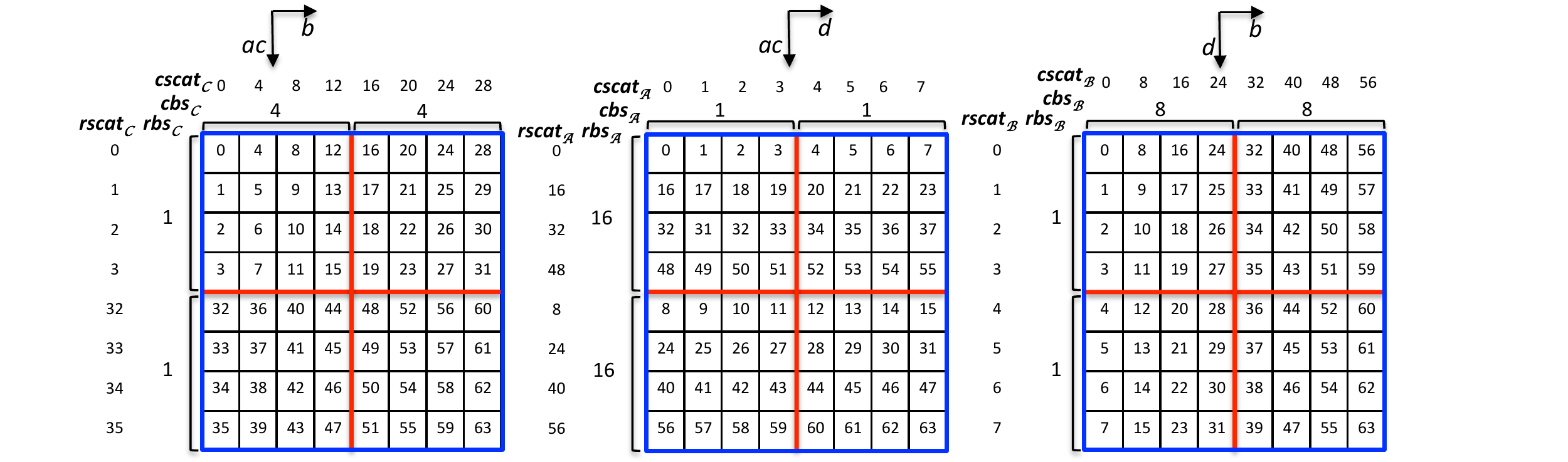}
\vspace{-0.2in}
\caption{Block scatter matrix view of \textbf{{(a)}}, where $\T{A}_{d,c,a}$, $\T{B}_{d,b}$, and $\T{C}_{a,b,c}$
are mapped to matrices $A_{i,p}$, $B_{p,j}$, and $C_{i,j}$:
\textbf{$rscat_{\T{T}}$} and \textbf{$cscat_{\T{T}}$} denote the scatter vectors; \textbf{$rbs_{\T{T}}$} and
\textbf{$cbs_{\T{T}}$} denote the block scatter vectors. Element locations
are given by the sum of the row and column scatter vector entries.
}
\label{fig:bstm}
\end{subfigure}
\caption{
An example to illustrate Strassen's algorithm for tensor contraction.
The red lines denotes \strassen{} $2\times2$ partitions mapping from block scatter matrix view (bottom) to the original tensor (top).
In this example the partitions are regular subtensors, but this is not required in general.
	}
\label{fig:stra_tensor}
\end{figure*}

In \figref{fig:tensor_contraction}, the tensor contraction $ \T{C}_{a,b,c} +\!\!= \T{A}_{d,c,a} \cdot \T{B}_{d,b} $ is illustrated.
In the general notation this gives
$I_m=\{a,c\}$, $J_m=\{b\}$, $P_k=\{d\}$,
$\pi_{\T{A}}(0,1,2)=\{2,1,0\}$,
$\pi_{\T{B}}(0,1)=\{0,1\}$,
and $\pi_{\T{C}}(0,1,2)=\{0,2,1\}$.
The number of floating point operations
and memory accesses 
for this contraction is identical to that for a
matrix multiplication of $(N_a \cdot N_c) \times N_d$, $N_d \times N_b$, and $(N_a \cdot N_c) \times N_b$ matrices.

\noindent
{\bf General stride layouts.}
The well-known column-major and row-major matrix layouts may be extended to tensors as the generalized
column- and row-major tensor layouts, where elements are stored contiguously along the first dimension or last
dimension, respectively. However, in general we may assume only a \emph{general tensor layout}, which extends the
general matrix layout \cite{BLIS1} by replacing matrix row and column strides ($rs_M$ and $cs_M$) with a stride
associated to each tensor dimension. For a $d$-dimensional tensor $\T{T}$ indexed by $T_d$, the strides
$s_{\T{T};k} \in \mathbb{N} $ for all $ 0 \le k < d$ form the set $S_{\T{T}} = \{s_{\T{T};0},\ldots,s_{\T{T},d-1}\}$,
which gives element \underline{LOC}ations relative to $\T{T}_{0,\ldots,0}$,
\[
LOC_{gsten}(\T{T}_{T_d}, S_{\T{T}}) = \sum\limits_{k=0}^{d-1} t_k \cdot s_{\T{A};k}.
\]
For convenience, we may also refer to the stride of the dimension indexed in $\T{T}$ by a particular symbol $x$ as $s_{\T{T};x}$.
The generalized column-major and row-major layouts can also be represented using a general stride layout, in which case
$s_{\T{T};k} = \prod_{l=0}^{k-1} N_{\T{T};l}$ and $s_{\T{T};k} = \prod_{l=k+1}^{d-1} N_{\T{T};l}$,
respectively.

In \figref{fig:tensor_contraction},
$\T{C}$ is stored in the generalized column-major layout.
The entries represents the location of the element
$\T{C}_{{a,b,c}}$ relative to the element $ \T{C}_{{0,0,0}}$
in the tensor storage layout.
$s_{\T{C};a} = 1$,
$s_{\T{C};b} = N_{\T{C};a} = 4$,
and $s_{\T{C};c} = N_{\T{C};a} \cdot N_{\T{C};b} = 32$.
The element location of $\T{C}_{{a,b,c}}$ is
$ a \cdot s_{\T{C};a} + b \cdot s_{\T{C};b} + c \cdot s_{\T{C};c} = a + 4b + 32c $. 

\noindent
{\bf Block Scatter Matrix View.}
In \cite{TC:Devin}
it is shown 
that tensors can be represented in a matrix-centric layout that allows for a simple
but efficient implementation of tensor contraction using the BLIS framework. The main idea of that work is that the
locations of tensor elements of $\T{T}$ can be described in a matrix format, the scatter matrix layout,
for some matrix $M$ very similarly to the general stride matrix layout,
\begin{equation}
LOC_{scatmat}(M_{i,j},rscat_{\T{T}},cscat_{\T{T}}) = rscat_{\T{T};i} + cscat_{\T{T};j},
\label{eqn:scatmat}
\end{equation}
where $rscat_{\T{T}} \in \mathbb{N}^{N_i}$ and $cscat_{\T{T}} \in \mathbb{N}^{N_j}$. If we define the index
bundle $I_p$ of size $p$ as the set of indices of $\T{T}$ that map to columns of $M$, and the index bundle
$J_q$ of size $q$ (such that $p+q=d_{\T{T}}$) as the set of indices that map
to rows of $M$, then by inspection of the general stride layout we can see that the scatter vector
$rscat_{\T{T}}$ with respect to $I_p$ is given by,
\begin{align*}
rscat_{\T{T};i} =& \sum\limits_{k=0}^{p-1} i_k \cdot s_{\T{T};i_k},\; i=\sum\limits_{k=0}^{p-1} i_k \cdot \prod\limits_{l=0}^{k-1} N_{i_l}, \,\forall\, \{i_0,\ldots,i_{p-1}\} \in N_{i_0}\times\ldots\times N_{i_{p-1}}; \\
\end{align*}
and similarly for $cscat_{\T{T}}$ with respect to $J_q$.


The relative location of $ \T{C}_{{a,b,c}} $ in \figref{fig:tensor_contraction}, or $ C_{{i,j}} $
in the matrix view of $ \T{C} $
in \figref{fig:bstm} is
$ rscat_{\T{C};i} + cscat_{\T{C};j} $ (e.g,
$ LOC( \T{C}_{2,3,1} ) = LOC( C_{6,3} )
=  rscat_{\T{C};6} + cscat_{\T{C};3} 
= 34 + 12 $).
Here:
(1) $rscat_{\T{C};i} = a \cdot s_{\T{C};a} + c \cdot s_{\T{C};c} = a + 32c,
i = a + c \cdot N_a = a + 4c,
\forall\, \{a,c\} \in N_a \times N_c$;
(2) $cscat_{\T{C};j} = b \cdot s_{\T{C};b} = 8b,
j = b,
\forall\, \{b\} \in N_b$.
These scatter vectors are shown on top and left of the matrix view of $\T{C}$ in \figref{fig:bstm}.

The general definition of tensor contractions gives a natural mapping from tensors to matrices through the index bundles
$I_m$, $J_n$, and $P_k$. Thus, the bundle $I_m$ defines $rscat_{\T{A}}$ and $rscat_{\T{C}}$, $J_n$ defines
$cscat_{\T{B}}$ and $cscat_{\T{C}}$, and $P_k$ defines $cscat_{\T{A}}$ and $rscat_{\T{B}}$. If we define matrices
$A_{i,k}$, $B_{k,j}$, and $C_{i,j}$ and imbue them with scatter matrix layouts using the scatter vectors from the corresponding
tensors, we can perform tensor contraction using
\jianyuFromTo{any suitable matrix multiplication algorithm.}{the high-performance matrix multiplication algorithm introduced in \secref{sec:gemm}, without
explicitly forming those matrices in extra working buffers.}

Since we are using the \jianyuFromTo{}{\GOTO{}/}BLIS algorithm, we can \jianyuFromTo{use}{leverage} the fact that these matrices will be partitioned to introduce further
optimizations.
\jianyuFromTo{At the lowest levels of the algorithm,}{In the \emph{micro-kernel} (\figref{fig:side_by_side}),}
the matrix $C$ will be partitioned into $m_R\times n_R$ blocks and the
matrices $A$ and $B$ will be partitioned into $m_R\times k_C$ and $k_C\times n_R$ slivers, respectively. If we further
partition $k_C$ into smaller increments of a new parameter \jianyuFromTo{$k_P$}{$k_R$}, on the order of $m_R$ and $n_R$, then we will end up
with only matrix blocks of very small size. As in \cite{TC:Devin}, we can partition the scatter vectors into very small blocks of
size $m_R$, $n_R$, and \jianyuFromTo{$k_P$}{$k_R$} as well, and use optimized algorithms in the \emph{packing} kernels and \jianyuFromTo{microkernel}{\emph{micro-kernel}} when the scatter
values for the current block are regularly spaced (i.e. strided). The regular strides for each \jianyuFromTo{$m_R$}{$m_R$/$n_R$/$k_R$}-sized block of
\jianyuFromTo{$rscat_{\T{T}}$}{$rscat_{\T{T}}$/$cscat_{\T{T}}$}, or zero if no regular stride exists, are collected in a \jianyuFromTo{row}{row/column} block scatter vector \jianyuFromTo{$rbs_{\T{T}}$}{$rbs_{\T{T}}$/$cbs_{\T{T}}$} of length
\jianyuFromTo{$\lceil \tfrac{N_i}{m_R} \rceil$}{$\lceil \tfrac{N_i}{m_R} \rceil$/$\lceil \tfrac{N_i}{n_R} \rceil$/$\lceil \tfrac{N_i}{k_R} \rceil$} and similarly for the other \jianyuFromTo{row}{row/column} scatter vectors.
With these block scatter vectors, we can then utilize efficient {\tt SIMD} vector load/store instructions for stride-one index,
or vector gather/scatter fetch instructions for stride-$n$ index, in a favorable memory access pattern.


In \figref{fig:bstm},
assuming $ m_R = n_R = k_R = 4 $,
$ rbs_{\T{C}}=\{1,1\} $,
and 
$ cbs_{\T{C}}=\{4,4\} $,
since the regular strides for each 4 elements of
$ rscat_{\T{C}} $ and $ cscat_{\T{C}} $ are 1 and 4, respectively.

\section{Strassen for Tensor Contraction}

The operations summarized in \secref{sec:strassen} are all special cases of
\begin{equation}
M = \alpha ( X + \delta Y)( V + \epsilon W); ~~~C +\!\!= \gamma_0 M; ~~~D+\!\!= \gamma_1 M;
\label{e:straprim1}
\end{equation}
for appropriately chosen $ \gamma_0, \gamma_1, \delta, \epsilon \in \{ -1, 0, 1 \} $.
Here, $X$ and $Y$ are submatrices of $A$, $V$ and $W$ are submatrices of $B$, and $C$ and $D$ are submatrices of the original $C$.
As in \cite{Strassen:SC16}, this scheme can be extended to multiple levels of \strassen{}.

Instead of partitioning the tensor $\T{A}$ into subtensors $\T{X}$ and $\T{Y}$ and so on for $\T{B}$ and $\T{C}$, we partition the matrix
representations $A$, $B$, and $C$ as in the matrix implementation of \strassen.\
\figref{fig:stra_tensor} provides an example to illustrate the partition mechanism.
Block scatter matrix layouts for these submatrices may be trivially obtained
by partitioning the scatter- and block scatter vectors of the entire matrices along the relevant dimensions. Once imbued with the appropriate
layouts, these submatrices may then be used in the BLIS-based \strassen\ of \cite{Strassen:SC16} along with modifications the the packing
kernels and micro-kernel as in \cite{TC:Devin}.

In fusing these two methodologies,
we need to further address 
the consideration of multiple block scatter vectors
as required when packing and executing the micro-kernel.
Methods for dealing with this issue are described in \secref{sec:packing}. The advantage of using
matrix partitions (which is enabled by the block scatter layout) instead of tensor partitions is primarily that only the product of the lengths of each
index bundle, $\{N_{I_m}$, $N_{J_n}$, $N_{P_k}\}$, must be considered when partitioning, and not the lengths of individual tensor dimensions. For
example, \strassen\ may be applied to any tensor contraction where \emph{at least} one dimension in each bundle is even in our approach, whereas
the $last$ dimension (or rather, the dimension with the longest stride) must be even when using subtensors.\footnote{A dimension other
than the last could also be chosen for partitioning, but the spatial locality of the partitioning would be destroyed.}
Additionally, when applying methods for performing \strassen\ on odd-length matrices to tensors, such as dynamical peeling as in \cite{Strassen:SC16}
or zero-padding, the overhead is larger for subtensors since a single dimension must be padded or peeled rather than the entire index bundle.

\section{Implementations}

The modifications to the block scatter matrix-based packing kernel and micro-kernel as described in \cite{TC:Devin} for \strassen\ are detailed.

\subsection{Packing}
\label{sec:packing}

When packing submatrices for \strassen\ using \eqref{e:straprim1}, multiple scatter- and block scatter vectors must be
considered. In our initial implementation, the block scatter vector entries for the corresponding block in both input submatrices (or all
submatrices for $L$-level \strassen) are examined. If \emph{all} entries are non-zero, then the constant stride is used in packing the
current block.\footnote{Note that when non-zero, the block scatter vector entries for different submatrices will always be equal.}
Otherwise, the scatter vectors are used when packing the current block, even though one or more of the input submatrix blocks
may in fact have a regular stride. In future work, we plan to exploit these cases for further performance improvements.

In addition to the \ABCstrassen\ algorithm, we also implement the \ABXstrassen\ and \XXXstrassen\ algorithms
of \cite{Strassen:SC16} for tensor contraction. In the \ABXstrassen\ algorithm, intermediate submatrices $M$ are explicitly stored
and then accumulated into submatrices of $C$. We store the $M$ submatrices as regular, densely-stored matrices, and handle their
accumulation onto block scatter matrix layout submatrices of $C$ using an adapted version of the \strassen\ block scatter matrix packing kernel.
In the \XXXstrassen\ algorithm, submatrices of $A$ and $B$ are also explicitly copied using a modified packing kernel and
stored as regular submatrices. Thus, the \XXXstrassen\ algorithm for tensor contraction is extremely similar to a \ttdt{}-based
\strassen\ algorithm (see \secref{sec:related}), except that the tensors are not required to be partitioned into regular subtensors.

\subsection{Micro-kernel}
\label{sec:microkernel}

As in \cite{Strassen:SC16}, we use assembly-coded micro-kernels that include the update to several submatrices of $C$ from
registers. In order to use this efficient update, $all$ block scatter vector entries for the relevant submatrix blocks of $C$ must be
non-zero. Unlike in the packing kernel implementation, the case where only one or more of the submatrix blocks is regular stride
would be more difficult to take advantage of, as the micro-kernel would have to be modified to flexibly omit or redirect individual
submatrix updates.

\section{Performance Model}
\label{sec:model}

In \cite{Strassen:SC16}, a performance model was proposed to predict the execution time $ T $ for variations of \strassen{} for matrices.
In this section, we extend that performance model to estimate the execution time $ T $ of ABC, AB and Naive variations of
$L$-level \strassen{} for TC and the high-performance TC routine we build on (see \secref{sec:tensor}; using TBLIS implementation \cite{TC:Devin,TBLIS_Software} introduced in \secref{sec:exp}; denoted as \tblis{} henceforth). 
Due to the high dimensionality of tensors and enormous types and combinations of permutations (transpositions) in TC,
it is impractical to exhaustively search for every tensor shape and tensor problem size to find the best variation.
Performance modeling helps us to better understand the memory footprint and computation of different \strassen{} implementations for TC, and at least reduce the
search space to pick the right implementation.
In our model, besides input problem size, block sizes, and the hardware parameters such as the peak GFLOPS and bandwidth,
$ T $ also depends on 
the shape of the tensors, and the extra permutations (transpositions) in the packing routines and in the micro-kernel.




\noindent
{\bf Notations.}
We summarize our notations in \figref{tab:notation}.
The total execution time, $ T $, can be decomposed of arithmetic time $ T_a $ and memory time $ T_m $ (\mycircle{2} in \figref{tab:perfmodel}).

\begin{figure}[!t]
\centering
{
\setlength{\tabcolsep}{2pt}
  \begin{tabular}{@{}c | l }
  \whline
  $ \tau_a $ & Time (in seconds) of one \underline{a}rithmetic (floating point) operation. \\
\hline
  \multirow{2}{*}{$ \tau_b $} & (\underline{B}andwidth) Amortized time (in seconds) of 8 Bytes contiguous\\
  & data movement from slow main memory to fast cache. \\
\hline
  $ \rho_a $ & Penalty factor for \underline{a}rithmetic operation effciency.\\
\hline
  $ \rho_b $ & Penalty factor for \underline{b}andwidth.\\
\hline
  $ T $     & Total execution time (in seconds). \\
\hline
  $ T_a $   & Time for \underline{a}rithmetic operations (in seconds). \\
\hline
  $ T_m $   & Time for \underline{m}emory operations (in seconds). \\
\hline
  $T_{a}^{\times}$ & $ T_a $ for (sub)tensor contractions. \\
\hline
  $T_{a}^{\T{A}_{+}}$, $T_{a}^{\T{B}_{+}}$, $T_{a}^{\T{C}_{+}}$ & $ T_a $ for extra (sub)tensor addtions/permutations. \\
\hline
  $T_{m}^{\T{A}_{\times}}$, $T_{m}^{\T{B}_{\times}}$ & $ T_m $ for reading (sub)tensors in packing routines (Fig. \ref{fig:side_by_side}).\\
\hline
  $T_{m}^{{\widetilde A}_{\times}}$,$T_{m}^{{\widetilde B}_{\times}}$  & $ T_m $ for writing into packed matrices in packing routines (Fig. \ref{fig:side_by_side}).\\
\hline
  $T_{m}^{\T{C}_{\times}}$ & $ T_m $ for reading \emph{and} writing (sub)tensors in  micro-kernel (Fig. \ref{fig:side_by_side}).\\
\hline
  \multirow{2}{*}{$T_{m}^{\T{A}_{+}}$, $T_{m}^{\T{B}_{+}}$, $T_{m}^{\T{C}_{+}}$} &
$ T_m $ for reading \emph{or} writing (sub)tensors, related to the \\
   & temporary buffer as part of \XXXstrassen{} and \ABXstrassen{}. \\
\hline
  $ \coeffa^{X}/\coeffm^{X} $ & Coefficient for the corresponding $ T_a^{X}/T_m^{X} $. \\
  \whline
  \end{tabular}
}
\caption{Notation table for performance model.}
\label{tab:notation}
\end{figure}

\begin{figure}[!t]
\centering
{
\setlength{\tabcolsep}{3pt}
  \begin{tabular}{l | l cr r }
  \whline
  \mycircle{1} & $\text{\emph{Effective} GFLOPS} = 2\cdot \NIm{}\cdot \NJn{}\cdot \NPk{} / T \cdot 10 ^{-9}$ \\
\hline
  \mycircle{2} & $T=T_{a}+T_{m}$ \\
\hline
  \mycircle{3} & $T_{a} = \coeffa^{\times} \cdot T_{a}^{\times} + \coeffa^{\T{A}_{+}} \cdot T_{a}^{\T{A}_{+}} + \coeffa^{\T{B}_{+}} \cdot T_{a}^{\T{B}_{+}} + \coeffa^{\T{C}_{+}} \cdot T_{a}^{\T{C}_{+}} $ \\
\hline
  \multirow{2}{*}{\mycircle{4}} & $T_{m} = \coeffm^{\T{A}_{\times}} \cdot T_{m}^{\T{A}_{\times}} + \coeffm^{\T{B}_{\times}} \cdot T_{m}^{\T{B}_{\times}} + \coeffm^{\T{C}_{\times}} \cdot T_{m}^{\T{C}_{\times}}$  \\
       & \hspace{0.4in}$+ \coeffm^{\T{A}_{+}} \cdot T_{m}^{\T{A}_{+}} + \coeffm^{\T{B}_{+}} \cdot T_{m}^{\T{B}_{+}} + \coeffm^{\T{C}_{+}} \cdot T_{m}^{\T{C}_{+}}$ \\
\hline
  \mycircle{5} & $\tau_{a} =  1 / ( \rho_{a} \cdot \mbox{Peak~GFLOPS} )$ \\
\hline
  \mycircle{6} &  $\tau_{b} = 8 / (\rho_{b} \cdot \mbox{Bandwidth} )$ \\
  \whline
  \end{tabular}
}

\vspace{0.05in}

\centering
{
  \begin{tabular}{l| ccr r }
  \whline
                                        & type
                                        & $\tau$
                                        & \tblis{}
                                        & $L$-level  \\
  \hline
  $T_{a}^{\times}$                      & -
                                        & $\tau_{a}$
                                        & $ 2 \NIm{}  \NJn{}  \NPk{} $
                                        & $ 2 \frac{\NIm{}}{\DivisorML{}}\frac{\NJn{}}{\DivisorNL}\frac{\NPk{}}{\DivisorKL}$ \\
  $T_{a}^{\T{A}_{+}}$                       & -
                                        & $\tau_{a}$
                                        & -
                                        & $ 2 \frac{\NIm{}}{\DivisorML{}}\frac{\NPk{}}{\DivisorKL} $ \\
  $T_{a}^{\T{B}_{+}}$                       & -
                                        & $\tau_{a}$
                                        & -
                                        & $ 2 \frac{\NPk{}}{\DivisorKL{}}\frac{\NJn{}}{\DivisorNL{}} $ \\
  $T_{a}^{\T{C}_{+}}$                       & -
                                        & $\tau_{a}$
                                        & -
                                        & $ 2 \frac{\NIm{}}{\DivisorML{}}\frac{\NJn{}}{\DivisorNL{}} $ \\
  \hline
  $T_{m}^{\T{A}_{\times}}$                  & \texttt{r}
                                        & $\tau_{b}$
                                        & $\NIm{}\NPk{} \lceil \frac{\NJn{}}{n_c} \rceil$
                                        & $\frac{\NIm{}}{ \DivisorML{}} \frac{\NPk{}}{ \DivisorKL{} } \lceil \frac{\NJn{}/\DivisorNL{}}{n_c} \rceil$ \\
  $T_{m}^{{\widetilde A}_{\times}}$     & \texttt{w}
                                        & $\tau_{b}$
                                        & $\NIm{}\NPk{} \lceil \frac{\NJn{}}{n_c} \rceil$
                                        & $\frac{\NIm{}}{\DivisorML{}} \frac{\NPk{}}{ \DivisorKL{}} \lceil \frac{\NJn{}/\DivisorNL{}}{n_c} \rceil$ \\
  $T_{m}^{\T{B}_{\times}}$                  & \texttt{r}
                                        & $\tau_{b}$
                                        & $\NJn{}\NPk{}$
                                        & $\frac{\NJn{}}{\DivisorNL{}} \frac{\NPk{}}{ \DivisorKL{}}$ \\
  $T_{m}^{{\widetilde B}_{\times}}$       & \texttt{w}
                                        & $\tau_{b}$
                                        & $\NJn{}\NPk{}$
                                        & $\frac{\NJn{}}{ \DivisorNL{}} \frac{\NPk{}}{\DivisorKL{}}$ \\
  $T_{m}^{\T{C}_{\times}}$ \fromto{(*)}{}   & \texttt{r/w}
                                        & $\tau_{b}$
                                        & $2\lambda \NIm{}\NJn{}\lceil\frac{\NPk{}}{k_c}\rceil$
                                        & $2\lambda \frac{\NIm{}}{\DivisorML{}}\frac{\NJn{}}{\DivisorNL{}}\lceil\frac{\NPk{}/\DivisorKL{}}{k_c}\rceil $ \\
  \hline
  $T_{m}^{\T{A}_{+}}$                       & \texttt{r/w}
                                        & $\tau_{b}$
                                        & $\NIm{}\NPk{}$
                                        & $\frac{\NIm{}}{\DivisorML{}} \frac{\NPk{}}{\DivisorKL{}}$ \\
  $T_{m}^{\T{B}_{+}}$                       & \texttt{r/w}
                                        & $\tau_{b}$
                                        & $\NJn{}\NPk{}$
                                        & $\frac{\NJn{}}{\DivisorNL{}} \frac{\NPk{}}{\DivisorKL{}}$ \\
  $T_{m}^{\T{C}_{+}}$                       & \texttt{r/w}
                                        & $\tau_{b}$
                                        & $\NIm{}\NJn{}$
                                        & $\frac{\NIm{}}{\DivisorML{}} \frac{\NJn{}}{\DivisorNL{}}$ \\
  \whline
  \end{tabular}
}

\vspace{0.05in}


\centering
{
\setlength{\tabcolsep}{3pt}
  \begin{tabular}{ c | c | c | c | c | c | c | c }
  \whline
                                        & \multirow{2}{*}{\tblis{}}    &   \multicolumn{3}{c |} { 1-level }  & \multicolumn{3}{c} { 2-level }  \\ \cline{3-8}
                                        &                              & ABC & AB & Naive                  & ABC & AB & Naive \\
  \hline
  $\coeffa^{\times}$                  & $ 1  $
                                        & $ 7  $ & $ 7  $ & $ 7  $ 
                                        & $ 49 $ & $ 49 $ & $ 49 $ \\
  $\coeffa^{\T{A}_{+}}$                   & -
                                        & $ 5  $ & $ 5  $ & $ 5  $
                                        & $ 95 $ & $ 95 $ & $ 95 $ \\
  $\coeffa^{\T{B}_{+}}$                   & -
                                        & $ 5  $ & $ 5  $ & $ 5  $
                                        & $ 95 $ & $ 95 $ & $ 95 $ \\
  $\coeffa^{C_{+}}$                   & -
                                        & $ 12  $ & $ 12  $ & $ 12  $
                                        & $ 144 $ & $ 144 $ & $ 144 $ \\
  \hline
  $\coeffm^{\T{A}_{\times}}$              & $ 1 $
                                        & $ 12  $ & $ 12  $ & $ 7   $
                                        & $ 194 $ & $ 194 $ & $ 49  $ \\
  $\coeffm^{{\widetilde A}_{\times}}$ & -
                                        & - & - & - 
                                        & - & - & - \\
  $\coeffm^{\T{B}_{\times}}$              & $ 1  $
                                        & $ 12  $ & $ 12  $ & $ 7   $
                                        & $ 194 $ & $ 194 $ & $ 49  $ \\
  $\coeffm^{{\widetilde B}_{\times}}$ & -
                                        & - & - & - 
                                        & - & - & - \\
  $\coeffm^{\T{C}_{\times}}$              & $ 1 $
                                        & $ 12  $ & $ 7   $ & $ 7   $
                                        & $ 144 $ & $ 49  $ & $ 49  $ \\
  \hline
  $\coeffm^{\T{A}_{+}}$                   & -
                                        & - & - & $ 19  $
                                        & - & - & $ 293 $ \\
  $\coeffm^{\T{B}_{+}}$                   & -
                                        & - & - & $ 19  $
                                        & - & - & $ 293 $ \\
  $\coeffm^{\T{C}_{+}}$                   & -
                                        & - & $ 36  $ & $ 36  $
                                        & - & $ 432 $ & $ 432 $ \\
  \whline
  \end{tabular}

}

\caption{
The top table shows
the equations for computing the execution time $ T $ and \emph{Effective} GFLOPS in our performance model.
The middle table shows the various components of arithmetic and memory operations for \tblis{} TC and various implementations of \strassen{} TC.
The time shown in the first column for \tblis{} TC and $L$-level \strassen{}
can be computed separately by multiplying the parameter in $\tau$ column with
the arithmetic/memory operation number in the corresponding entries.
The bottom table shows
the coefficient $\coeffa^X$/$\coeffm^X$ mapping table for computing $T_a^{X}$/$T_m^{X}$ in the performance model.
Here
$ \NIm{} = \prod_{i \in I_m} N_i = N_{i_0} \cdot \ldots \cdot N_{i_{m-1}} $,
$ \NJn{} = \prod_{j \in J_n} N_j = N_{j_0} \cdot \ldots \cdot N_{j_{n-1}} $,
$ \NPk{} = \prod_{p \in P_k} N_p = N_{p_0} \cdot \ldots \cdot N_{p_{k-1}} $.
}
\label{tab:perfmodel}
\end{figure}

\noindent
{\bf Arithmetic operations.}
As shown in \mycircle{3}, $ T_a $ includes (sub)tensor contraction ($T_{a}^{\times}$) and (sub)tensor additions/permutations ($T_{a}^{\T{A}_{+}}$, $T_{a}^{\T{B}_{+}}$, $T_{a}^{\T{C}_{+}}$). The corresponding coefficients $W_{a}^{X}$ for \tblis{} TC and $L$-level various \strassen{} TC are enumerated in \figref{tab:perfmodel}. Note that $T_{a}^{X}$ is calculated by multiplying the unit time $ \tau_a $ with the arithmetic operation number in the middle table of \figref{tab:perfmodel}.
We compute $ \tau_a $ through \mycircle{5}.
The penalty factor $ \rho_a \in (0,1]$ is introduced, due to the extra computations involved in 
$rscat_{\T{T}}$/$cscat_{\T{T}}$/$rbs_{\T{T}}$/$cbs_{\T{T}}$,
and the slow micro-kernel invocation when the corresponding entries in $rbs_{\T{C}}$ or $cbs_{\T{C}}$ are $0$ (see \secref{sec:microkernel}; non-regular stride access).
We penalize the performance drops caused by these factors by setting
$ \rho_a = 0.95 $.

%





\noindent
{\bf Memory operations.}
Similar to \cite{Strassen:SC16}, we assume two layers of modern memory hierarchy: slow main memory and fast caches.
For write operations, the lazy write-back policy is enforced such that the time for writing into fast caches can be hidden.
For read operations, the latency for accessing the slow main memory is counted, while the latency for accessing caches can be ignored.
With these assumptions, $ T_m $ can be broken down into three parts (\mycircle{4} in \figref{tab:perfmodel}):
updating the temporary buffer that are parts of \XXXstrassen{}/\ABXstrassen{} ($\coeffm^{\T{T}_{+}} \cdot T_{m}^{\T{T}_{+}}$);
memory packing shown in \figref{fig:side_by_side}
($\coeffm^{\T{A}_{\times}} \cdot T_{m}^{\T{A}_{\times}}$
, $\coeffm^{\T{B}_{\times}} \cdot T_{m}^{\T{B}_{\times}}$)
;
updating the submatrices of $ C $ shown in \figref{fig:side_by_side}
($\coeffm^{\T{C}_{\times}} \cdot T_{m}^{\T{C}_{\times}}$).
The coefficients $ \coeffm^X $ are tabulated in \figref{tab:perfmodel}.
$T_{m}^X$ is a function of block sizes $ \{m_C, k_C, n_C\} $ in \figref{fig:side_by_side}, and the bundle lengths $\{ \NIm{}/2^L, \NJn{}/2^L, \NPk{}/2^L \}$
because the memory operation can repeat multiple times according to which loop they reside in.
\figref{tab:perfmodel}(middle) characterizes each memory operation term by its read/write type and the amount of memory in units of 64-bit double precision elements.
$T_m^{{\widetilde A}_{\times}}$, 
$T_m^{{\widetilde B}_{\times}}$
are omitted in \mycircle{4} due to the lazy write-back policy assumption.
Because of the software prefetching effects,
there is an extra parameter $\lambda \in (0.5,1]$ for $T_m^{C_{\times}}$, which denotes the prefetching efficiency.
In order to get $T_{m}^X$, the memory operation number needs to be multiplied by the bandwidth $ \tau_b $.
We compute $ \tau_b $ through \mycircle{6}.
We penalize the effect of permutations without stride-one index accesss (see \secref{sec:packing}; the corresponding entries in neither $rbs_{\T{T}}$ or $cbs_{\T{T}}$ are 1, i.e. using scatter/gather operation, or indirect memory addressing with (\ref{eqn:scatmat}))
by setting $ \rho_b = 0.7 $.
A similar parameter is introduced in \cite{Paul16} for regular TC.

\begin{figure*}[htp!]
\center
\begin{tikzpicture}[scale=0.9]
\begin{axis}[
    title={$\NImCap{} = \NPkCap{} = \NJnCap{}$, 1 core, \emph{Actual vs. Modeled}},
    xlabel={ $\NImCap{} = \NPkCap{} = \NJnCap{}$ },
    ylabel={\emph{Effective} GFLOPS ($ 2 \cdot \NIm{} \cdot \NJn{} \cdot \NPk{} / time $)},
    xmin=0,
    xmax=12000,
    ymin=8,
    ymax=33,
    xtick={1000,2000,3000,4000,5000,6000,7000,8000,9000,10000,11000,12000},
    ytick={5, 10, 15, 20, 25, 28.36, 32},
    scaled x ticks=false,
    scaled x ticks=base 10:-3,
    grid=major,
    axis background/.style={fill=lightgray!20},
    mark size=0.8pt,
    cycle list name=jianyucolorboth,
    restrict y to domain=1:inf,
    legend style={
        at={(0.99,0.01)},
        anchor=south east,
        legend columns=7,
        transpose legend,
        font=\tiny,
        rounded corners=1pt,
        nodes={scale=0.75, transform shape},
        cells={anchor=west},
    },
    legend entries = {
        $ TBLIS             ~(Actual)  $\\
        $ One$-$level~ABC   ~(Actual)  $\\ 
        $ One$-$level~AB    ~(Actual)  $\\ 
        $ One$-$level~Naive ~(Actual)  $\\ 
        $ Two$-$level~ABC   ~(Actual)  $\\ 
        $ Two$-$level~AB    ~(Actual)  $\\ 
        $ Two$-$level~Naive ~(Actual)  $\\ 
        $ TBLIS             ~(Modeled) $\\
        $ One$-$level~ABC   ~(Modeled) $\\ 
        $ One$-$level~AB    ~(Modeled) $\\ 
        $ One$-$level~Naive ~(Modeled) $\\ 
        $ Two$-$level~ABC   ~(Modeled) $\\ 
        $ Two$-$level~AB    ~(Modeled) $\\ 
        $ Two$-$level~Naive ~(Modeled) $\\ 
     },
    ]
\addplot table[x=dim,y=tblis,col sep=comma] {plotdata/square_1core.csv};
\addplot table[x=dim,y=tensor_1_abc,col sep=comma] {plotdata/square_1core.csv};
\addplot table[x=dim,y=tensor_1_ab,col sep=comma] {plotdata/square_1core.csv};
\addplot table[x=dim,y=tensor_1_naive,col sep=comma] {plotdata/square_1core.csv};
\addplot table[x=dim,y=tensor_2_abc,col sep=comma] {plotdata/square_1core.csv};
\addplot table[x=dim,y=tensor_2_ab,col sep=comma] {plotdata/square_1core.csv};
\addplot table[x=dim,y=tensor_2_naive,col sep=comma] {plotdata/square_1core.csv};
\addplot table[x=dim,y=gemm,col sep=comma] {plotdata/model_square_1core.csv};
\addplot table[x=dim,y=tensor_1_abc,col sep=comma] {plotdata/model_square_1core.csv};
\addplot table[x=dim,y=tensor_1_ab,col sep=comma] {plotdata/model_square_1core.csv};
\addplot table[x=dim,y=tensor_1_naive,col sep=comma] {plotdata/model_square_1core.csv};
\addplot table[x=dim,y=tensor_2_abc,col sep=comma] {plotdata/model_square_1core.csv};
\addplot table[x=dim,y=tensor_2_ab,col sep=comma] {plotdata/model_square_1core.csv};
\addplot table[x=dim,y=tensor_2_naive,col sep=comma] {plotdata/model_square_1core.csv};
\end{axis}
\end{tikzpicture}
\begin{tikzpicture}[scale=0.9]
\begin{axis}[
    title={$\NImCap{} = \NPkCap{} = \NJnCap{}$, 10 core, \emph{Actual}},
    xlabel={ $\NImCap{} = \NPkCap{} = \NJnCap{}$ },
    ylabel={\emph{Effective} GFLOPS ($ 2 \cdot  \NIm{} \cdot  \NJn{} \cdot \NPk{} / time $)},
    xmin=0,
    xmax=12000,
    ymin=60,
    ymax=255,
    xtick={1000,2000,3000,4000,5000,6000,7000,8000,9000,10000,11000,12000},
    ytick={50, 100, 150, 200, 248},
    scaled x ticks=false,
    scaled x ticks=base 10:-3,
    grid=major,
    axis background/.style={fill=lightgray!20},
    mark size=0.8pt,
    cycle list name=jianyucolor,
    restrict y to domain=1:inf,
    legend style={
        at={(0.99,0.01)},
        anchor=south east,
        legend columns=1,
        font=\tiny,
        rounded corners=1pt,
        nodes={scale=0.75, transform shape},
        cells={anchor=west},
    },
    legend entries = {
        $ TBLIS            $\\
        $ One$-$level~ABC  $\\ 
        $ One$-$level~AB   $\\ 
        $ One$-$level~Naive$\\ 
        $ Two$-$level~ABC  $\\ 
        $ Two$-$level~AB   $\\ 
        $ Two$-$level~Naive$\\ 
    },
    ]
\addplot table[x=dim,y=tblis,col sep=comma] {plotdata/square_10core.csv};
\addplot table[x=dim,y=tensor_1_abc,col sep=comma] {plotdata/square_10core.csv};
\addplot table[x=dim,y=tensor_1_ab,col sep=comma] {plotdata/square_10core.csv};
\addplot table[x=dim,y=tensor_1_naive,col sep=comma] {plotdata/square_10core.csv};
\addplot table[x=dim,y=tensor_2_abc,col sep=comma] {plotdata/square_10core.csv};
\addplot table[x=dim,y=tensor_2_ab,col sep=comma] {plotdata/square_10core.csv};
\addplot table[x=dim,y=tensor_2_naive,col sep=comma] {plotdata/square_10core.csv};
\end{axis}
\end{tikzpicture}
\begin{tikzpicture}[scale=0.9]
\begin{axis}[
    title={$\NImCap{} = \NJnCap{} = 16000$, 1 core, \emph{Actual vs. Modeled}},
    xlabel={ $\NPkCap{}$ },
    ylabel={\emph{Effective} GFLOPS ($ 2 \cdot \NIm{} \cdot \NJn{} \cdot \NPk{} / time $)},
    xmin=0,
    xmax=12000,
    ymin=8,
    ymax=33,
    xtick={1000,2000,3000,4000,5000,6000,7000,8000,9000,10000,11000,12000},
    ytick={5, 10, 15, 20, 25, 28.36, 32},
    scaled x ticks=false,
    scaled x ticks=base 10:-3,
    grid=major,
    axis background/.style={fill=lightgray!20},
    mark size=0.8pt,
    cycle list name=jianyucolorboth,
    restrict y to domain=1:inf,
    legend style={
        at={(0.99,0.01)},
        anchor=south east,
        legend columns=7,
        transpose legend,
        font=\tiny,
        rounded corners=1pt,
        nodes={scale=0.75, transform shape},
        cells={anchor=west},
    },
    legend entries = {
        $ TBLIS             ~(Actual)  $\\
        $ One$-$level~ABC   ~(Actual)  $\\ 
        $ One$-$level~AB    ~(Actual)  $\\ 
        $ One$-$level~Naive ~(Actual)  $\\ 
        $ Two$-$level~ABC   ~(Actual)  $\\ 
        $ Two$-$level~AB    ~(Actual)  $\\ 
        $ Two$-$level~Naive ~(Actual)  $\\ 
        $ TBLIS             ~(Modeled) $\\
        $ One$-$level~ABC   ~(Modeled) $\\ 
        $ One$-$level~AB    ~(Modeled) $\\ 
        $ One$-$level~Naive ~(Modeled) $\\ 
        $ Two$-$level~ABC   ~(Modeled) $\\ 
        $ Two$-$level~AB    ~(Modeled) $\\ 
        $ Two$-$level~Naive ~(Modeled) $\\ 
     },
    ]
\addplot table[x=dim,y=tblis,col sep=comma] {plotdata/rankk_1core.csv};
\addplot table[x=dim,y=tensor_1_abc,col sep=comma] {plotdata/rankk_1core.csv};
\addplot table[x=dim,y=tensor_1_ab,col sep=comma] {plotdata/rankk_1core.csv};
\addplot table[x=dim,y=tensor_1_naive,col sep=comma] {plotdata/rankk_1core.csv};
\addplot table[x=dim,y=tensor_2_abc,col sep=comma] {plotdata/rankk_1core.csv};
\addplot table[x=dim,y=tensor_2_ab,col sep=comma] {plotdata/rankk_1core.csv};
\addplot table[x=dim,y=tensor_2_naive,col sep=comma] {plotdata/rankk_1core.csv};
\addplot table[x=dim,y=gemm,col sep=comma] {plotdata/model_rankk_1core.csv};
\addplot table[x=dim,y=tensor_1_abc,col sep=comma] {plotdata/model_rankk_1core.csv};
\addplot table[x=dim,y=tensor_1_ab,col sep=comma] {plotdata/model_rankk_1core.csv};
\addplot table[x=dim,y=tensor_1_naive,col sep=comma] {plotdata/model_rankk_1core.csv};
\addplot table[x=dim,y=tensor_2_abc,col sep=comma] {plotdata/model_rankk_1core.csv};
\addplot table[x=dim,y=tensor_2_ab,col sep=comma] {plotdata/model_rankk_1core.csv};
\addplot table[x=dim,y=tensor_2_naive,col sep=comma] {plotdata/model_rankk_1core.csv};
\end{axis}
\end{tikzpicture}
\begin{tikzpicture}[scale=0.9]
\begin{axis}[
    title={$\NImCap{} = \NJnCap{} = 16000$, 10 core, \emph{Actual}},
    xlabel={ $\NPkCap{}$ },
    ylabel={\emph{Effective} GFLOPS ($ 2 \cdot  \NIm{} \cdot  \NJn{} \cdot \NPk{} / time $)},
    xmin=0,
    xmax=12000,
    ymin=60,
    ymax=255,
    xtick={1000,2000,3000,4000,5000,6000,7000,8000,9000,10000,11000,12000},
    ytick={50, 100, 150, 200, 248},
    scaled x ticks=false,
    scaled x ticks=base 10:-3,
    grid=major,
    axis background/.style={fill=lightgray!20},
    mark size=0.8pt,
    cycle list name=jianyucolor,
    restrict y to domain=1:inf,
    legend style={
        at={(0.99,0.01)},
        anchor=south east,
        legend columns=1,
        font=\tiny,
        rounded corners=1pt,
        nodes={scale=0.75, transform shape},
        cells={anchor=west},
    },
    legend entries = {
        $ TBLIS            $\\
        $ One$-$level~ABC  $\\ 
        $ One$-$level~AB   $\\ 
        $ One$-$level~Naive$\\ 
        $ Two$-$level~ABC  $\\ 
        $ Two$-$level~AB   $\\ 
        $ Two$-$level~Naive$\\ 
    },
    ]
\addplot table[x=dim,y=tblis,col sep=comma] {plotdata/rankk_10core.csv};
\addplot table[x=dim,y=tensor_1_abc,col sep=comma] {plotdata/rankk_10core.csv};
\addplot table[x=dim,y=tensor_1_ab,col sep=comma] {plotdata/rankk_10core.csv};
\addplot table[x=dim,y=tensor_1_naive,col sep=comma] {plotdata/rankk_10core.csv};
\addplot table[x=dim,y=tensor_2_abc,col sep=comma] {plotdata/rankk_10core.csv};
\addplot table[x=dim,y=tensor_2_ab,col sep=comma] {plotdata/rankk_10core.csv};
\addplot table[x=dim,y=tensor_2_naive,col sep=comma] {plotdata/rankk_10core.csv};
\end{axis}
\end{tikzpicture}
\begin{tikzpicture}[scale=0.9]
\begin{axis}[
    title={$\NPkCap{} = 1024$, 1 core, \emph{Actual vs. Modeled}},
    xlabel={ $\NImCap{} = \NJnCap{}$ },
    ylabel={\emph{Effective} GFLOPS ($ 2 \cdot \NIm{} \cdot \NJn{} \cdot \NPk{} / time $)},
    xmin=0,
    xmax=12000,
    ymin=8,
    ymax=33,
    xtick={1000,2000,3000,4000,5000,6000,7000,8000,9000,10000,11000,12000},
    ytick={5, 10, 15, 20, 25, 28.36, 32},
    scaled x ticks=false,
    scaled x ticks=base 10:-3,
    grid=major,
    axis background/.style={fill=lightgray!20},
    mark size=0.8pt,
    cycle list name=jianyucolorboth,
    restrict y to domain=1:inf,
    legend style={
        at={(0.99,0.01)},
        anchor=south east,
        legend columns=7,
        transpose legend,
        font=\tiny,
        rounded corners=1pt,
        nodes={scale=0.75, transform shape},
        cells={anchor=west},
    },
    legend entries = {
        $ TBLIS             ~(Actual)  $\\
        $ One$-$level~ABC   ~(Actual)  $\\ 
        $ One$-$level~AB    ~(Actual)  $\\ 
        $ One$-$level~Naive ~(Actual)  $\\ 
        $ Two$-$level~ABC   ~(Actual)  $\\ 
        $ Two$-$level~AB    ~(Actual)  $\\ 
        $ Two$-$level~Naive ~(Actual)  $\\ 
        $ TBLIS             ~(Modeled) $\\
        $ One$-$level~ABC   ~(Modeled) $\\ 
        $ One$-$level~AB    ~(Modeled) $\\ 
        $ One$-$level~Naive ~(Modeled) $\\ 
        $ Two$-$level~ABC   ~(Modeled) $\\ 
        $ Two$-$level~AB    ~(Modeled) $\\ 
        $ Two$-$level~Naive ~(Modeled) $\\ 
     },
    ]
\addplot table[x=dim,y=tblis,col sep=comma] {plotdata/fixk_1core.csv};
\addplot table[x=dim,y=tensor_1_abc,col sep=comma] {plotdata/fixk_1core.csv};
\addplot table[x=dim,y=tensor_1_ab,col sep=comma] {plotdata/fixk_1core.csv};
\addplot table[x=dim,y=tensor_1_naive,col sep=comma] {plotdata/fixk_1core.csv};
\addplot table[x=dim,y=tensor_2_abc,col sep=comma] {plotdata/fixk_1core.csv};
\addplot table[x=dim,y=tensor_2_ab,col sep=comma] {plotdata/fixk_1core.csv};
\addplot table[x=dim,y=tensor_2_naive,col sep=comma] {plotdata/fixk_1core.csv};
\addplot table[x=dim,y=gemm,col sep=comma] {plotdata/model_fixk_1core.csv};
\addplot table[x=dim,y=tensor_1_abc,col sep=comma] {plotdata/model_fixk_1core.csv};
\addplot table[x=dim,y=tensor_1_ab,col sep=comma] {plotdata/model_fixk_1core.csv};
\addplot table[x=dim,y=tensor_1_naive,col sep=comma] {plotdata/model_fixk_1core.csv};
\addplot table[x=dim,y=tensor_2_abc,col sep=comma] {plotdata/model_fixk_1core.csv};
\addplot table[x=dim,y=tensor_2_ab,col sep=comma] {plotdata/model_fixk_1core.csv};
\addplot table[x=dim,y=tensor_2_naive,col sep=comma] {plotdata/model_fixk_1core.csv};
\end{axis}
\end{tikzpicture}
\begin{tikzpicture}[scale=0.9]
\begin{axis}[
    title={$\NPkCap{} = 1024$, 10 core, \emph{Actual}},
    xlabel={ $\NImCap{} = \NJnCap{}$ },
    ylabel={\emph{Effective} GFLOPS ($ 2 \cdot  \NIm{} \cdot  \NJn{} \cdot \NPk{} / time $)},
    xmin=0,
    xmax=12000,
    ymin=60,
    ymax=255,
    xtick={1000,2000,3000,4000,5000,6000,7000,8000,9000,10000,11000,12000},
    ytick={50, 100, 150, 200, 248},
    scaled x ticks=false,
    scaled x ticks=base 10:-3,
    grid=major,
    axis background/.style={fill=lightgray!20},
    mark size=0.8pt,
    cycle list name=jianyucolor,
    restrict y to domain=1:inf,
    legend style={
        at={(0.99,0.01)},
        anchor=south east,
        legend columns=1,
        font=\tiny,
        rounded corners=1pt,
        nodes={scale=0.75, transform shape},
        cells={anchor=west},
    },
    legend entries = {
        $ TBLIS            $\\
        $ One$-$level~ABC  $\\ 
        $ One$-$level~AB   $\\ 
        $ One$-$level~Naive$\\ 
        $ Two$-$level~ABC  $\\ 
        $ Two$-$level~AB   $\\ 
        $ Two$-$level~Naive$\\ 
    },
    ]
\addplot table[x=dim,y=tblis,col sep=comma] {plotdata/fixk_10core.csv};
\addplot table[x=dim,y=tensor_1_abc,col sep=comma] {plotdata/fixk_10core.csv};
\addplot table[x=dim,y=tensor_1_ab,col sep=comma] {plotdata/fixk_10core.csv};
\addplot table[x=dim,y=tensor_1_naive,col sep=comma] {plotdata/fixk_10core.csv};
\addplot table[x=dim,y=tensor_2_abc,col sep=comma] {plotdata/fixk_10core.csv};
\addplot table[x=dim,y=tensor_2_ab,col sep=comma] {plotdata/fixk_10core.csv};
\addplot table[x=dim,y=tensor_2_naive,col sep=comma] {plotdata/fixk_10core.csv};
\end{axis}
\end{tikzpicture}
\caption{Performance of various implementations for synthetic data on single core and one socket.
Left column: actual and modeled performance on single core; Right column: actual performance on one socket.
Top row: \SQUARE{}; Middle row: \RANKK{}; Bottom row: \FIXK{}.
}
\label{fig:perf_all}
\end{figure*}

\noindent
{\bf Discussion}
We can estimate the run time performance of various implementations, based on the performance model presented in \figref{tab:perfmodel}.
Here we define \emph{Effective} GFLOPS (\mycircle{1} in \figref{tab:perfmodel}) for TC as the metric to compare
the performance of various \strassen{} TC and \tblis{} TC.
The theoretical peak GFLOPS and bandwidth information is given in \secref{sec:exp}.
In \figref{fig:perf_all}(left), we demonstrate the modeled and actual performance for a wide range of synthetic tensor sizes and shapes: 
\SQUARE{}; \RANKK{}; \FIXK{}.
How we generate synthetic data is  detailed in \secref{sec:exp}.
\begin{itemize}[leftmargin=*]
\item For \SQUARE{},
the \ABCstrassen{}/\ABXstrassen{} implementations outperform \tblis{},
when $\NIm{}$, $\NJn{}$, $\NPk{}$ are as small as $ 2 k_C $, nearly 500;
while \XXXstrassen{} cannot beat \tblis{} until the problem size is larger than 2000.
\item The ``\RANKK{}'' graph shows that when $ \NPk{} $ is small, \ABCstrassen{} performs best; when $ \NPk{} $ is large, \ABXstrassen{} performs better.
The coefficients $\coeffm^X$ in \figref{tab:perfmodel}(bottom) help to illustrate the reasons quantitatively.
\item According to the model, when $ \NPk{} $ is equal to appropriate multiple of $ k_C $ ($ \NPk{} = 2^L \cdot k_C $ for $L$-level), \ABCstrassen{} achieves the best performance. We will leverage this observation in our distributed memory experiment.
\end{itemize}

\section{Experiments}
\label{sec:exp}

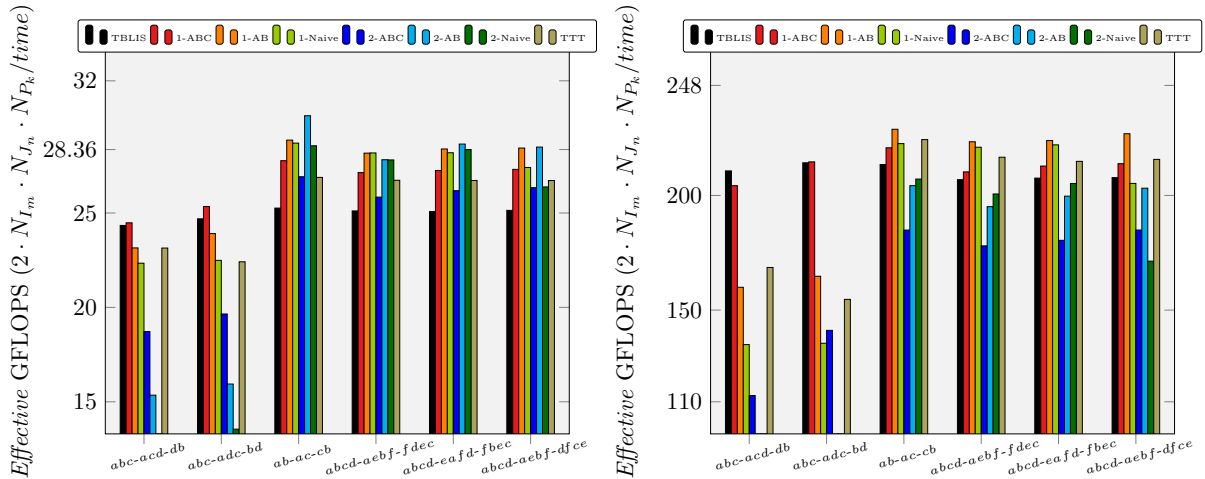
\begin{figure*}[htp!]
\center
\begin{tikzpicture}[scale=0.9]
\begin{axis}[
    ylabel={\emph{Effective} GFLOPS ($ 2 \cdot  \NIm{} \cdot  \NJn{} \cdot \NPk{} / time $)},
    enlargelimits=0.1,
    ymin=15,
    ymax=32,
    ybar=0*\pgflinewidth,
    ytick={10, 15, 20, 25, 28.36, 32},
    bar width=2.5pt,
    axis background/.style={fill=lightgray!20},
    xticklabels={0,$abc$-$dca$-$bd$,$abc$-$acd$-$db$,$abc$-$adc$-$bd$,$ab$-$ac$-$cb$,$abcd$-$aebf$-$fdec$,$abcd$-$eafd$-$fbec$,$abcd$-$aebf$-$dfce$,10},
    x tick label style={rotate=15,anchor=north,font=\tiny,align=center},
    legend style={
        at={(0.5,0.99)},
        anchor=south,
        legend columns=-1,
        font=\tiny,
        rounded corners=1pt,
        nodes={scale=0.75, transform shape},
        cells={anchor=west},
    },
    legend entries = {
        TBLIS  \\
        1-ABC  \\ 
        1-AB   \\ 
        1-Naive\\ 
        2-ABC  \\ 
        2-AB   \\ 
        2-Naive\\ 
        TTT\\ 
    },
    cycle list name=jianyucolorbar,
]

\addplot table[x=dim,y=tblis,col sep=comma] {plotdata/benchmark.csv};


\addplot table[x=dim,y=tensor_1_abc,col sep=comma] {plotdata/benchmark.csv};
\addplot table[x=dim,y=tensor_1_ab,col sep=comma] {plotdata/benchmark.csv};
\addplot table[x=dim,y=tensor_1_naive,col sep=comma] {plotdata/benchmark.csv};
\addplot table[x=dim,y=tensor_2_abc,col sep=comma] {plotdata/benchmark.csv};
\addplot table[x=dim,y=tensor_2_ab,col sep=comma] {plotdata/benchmark.csv};
\addplot table[x=dim,y=tensor_2_naive,col sep=comma] {plotdata/benchmark.csv};
\addplot table[x=dim,y=ttt,col sep=comma] {plotdata/benchmark.csv};


\end{axis}
\end{tikzpicture}
\begin{tikzpicture}[scale=0.9]
\begin{axis}[
    ylabel={\emph{Effective} GFLOPS ($ 2 \cdot  \NIm{} \cdot  \NJn{} \cdot \NPk{} / time $)},
    enlargelimits=0.1,
    ymin=110,
    ymax=250,
    ybar=0*\pgflinewidth,
    ytick={50, 110, 150, 200, 248},
    bar width=2.5pt,
    axis background/.style={fill=lightgray!20},
    xticklabels={0,$abc$-$dca$-$bd$,$abc$-$acd$-$db$,$abc$-$adc$-$bd$,$ab$-$ac$-$cb$,$abcd$-$aebf$-$fdec$,$abcd$-$eafd$-$fbec$,$abcd$-$aebf$-$dfce$,10},
    x tick label style={rotate=15,anchor=north,font=\tiny,align=center},
    legend style={
        at={(0.5,0.99)},
        anchor=south,
        legend columns=-1,
        font=\tiny,
        rounded corners=1pt,
        nodes={scale=0.75, transform shape},
        cells={anchor=west},
    },
    legend entries = {
        TBLIS  \\
        1-ABC  \\ 
        1-AB   \\ 
        1-Naive\\ 
        2-ABC  \\ 
        2-AB   \\ 
        2-Naive\\ 
        TTT\\ 
    },
    cycle list name=jianyucolorbar,
]

\addplot table[x=dim,y=tblis,col sep=comma] {plotdata/benchmark_10core.csv};


\addplot table[x=dim,y=tensor_1_abc,col sep=comma] {plotdata/benchmark_10core.csv};
\addplot table[x=dim,y=tensor_1_ab,col sep=comma] {plotdata/benchmark_10core.csv};
\addplot table[x=dim,y=tensor_1_naive,col sep=comma] {plotdata/benchmark_10core.csv};
\addplot table[x=dim,y=tensor_2_abc,col sep=comma] {plotdata/benchmark_10core.csv};
\addplot table[x=dim,y=tensor_2_ab,col sep=comma] {plotdata/benchmark_10core.csv};
\addplot table[x=dim,y=tensor_2_naive,col sep=comma] {plotdata/benchmark_10core.csv};
\addplot table[x=dim,y=ttt,col sep=comma] {plotdata/benchmark_10core.csv};


\end{axis}
\end{tikzpicture}
\caption{Performance for representative user cases of benchmark from \cite{Paul16}.
TC is identified by the index string, with the tensor index bundle of each tensor
in the order $\T{C}$-$\T{A}$-$\T{B}$,
e.g. 
$\T{C}_{abcd} := \T{A}_{aebf} \T{B}_{dfce}$ is denoted as $abcd$-$aebf$-$dfce$.
Left: performance on single core. Right: performance on one socket.
}
\label{fig:benchmark}
\end{figure*}


We perform our experimental evaluations for synthetic data and real-world benchmarks on a single node and on a distributed memory architecture.
The
implementations
are written in {\tt C++}, utilizing {\tt AVX} assembly,
based on the open source TBLIS
framework \cite{TBLIS_Software}.
We compare against TBLIS's tensor contraction routine (marked as \tblis{}) as well as the TTT routine from MATLAB Tensor Toolbox \cite{TTB_Software}
(linked with Intel MKL \cite{IntelMKL}, marked as \ttt{}) for single node, and tensor contraction routine from the Cyclops Tensor Framework \cite{CTF2014} (also linked with Intel MKL, marked with \ctf{}) for distributed memory.


We measure the CPU performance results on the Maverick system at the Texas Advanced Computing Center (TACC).  Each node of that system consists of a  
dual-socket (10 cores/socket) Intel Xeon E5-2680 v2 (Ivy Bridge) processors with 256 GB memory
(peak bandwidth: 59.7 GB/s with four channels) and a three-level cache (32 KB L1 data; 256 KB L2; 25.6 MB L3).
The stable CPU clockrate is 3.54 GHz when a single core is utilized (28.32 GFLOPS peak, marked in the graphs) and 3.10 GHz when all ten cores are in use (24.8 GFLOPS/core peak).
We disable hyper-threading explicitly and set thread affinity with {\tt KMP\_AFFINITY=compact} which also ensures the computation and the memory allocation all reside on 
the same socket. 

The cache blocking parameters, $m_C=96$, $n_C=4096$, $k_C=256$, 
and the register block sizes, $m_R=8$, $n_R=4$, 
are consistent with parameters used for the standard \BLIS\ \dgemm\ implementation for this architecture. We use the default
value of $k_R=4$ as defined in TBLIS.
This makes the size of the packing buffer
$ \widetilde A_i $ 192 KB and 
$ \widetilde B_p $ 8192 KB, which then fit the L2 cache and L3 cache, respectively. Parallelization 
is implemented mirroring that described in~\cite{BLIS3}, but with the number of threads assigned to each of
the loops in \figref{fig:side_by_side} automatically determined by the TBLIS framework.


\subsection{Single node experiments}

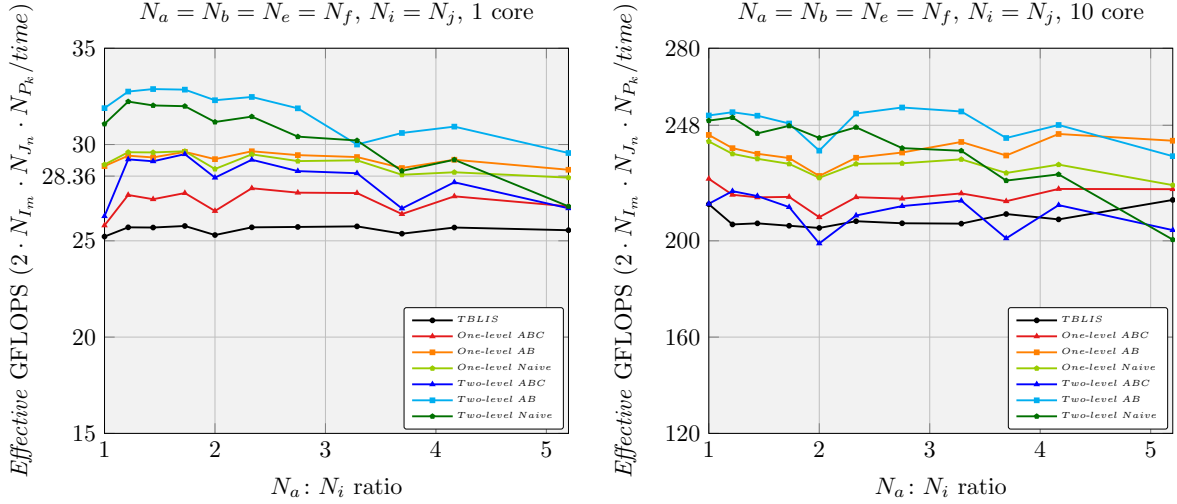
\begin{figure*}[htp!]
\center
\begin{tikzpicture}[scale=0.9]
\begin{axis}[
    title={$ N_a = N_b = N_e = N_f $, $ N_i = N_j $, 1 core},
    xlabel={ $N_a \colon N_i$ ratio },
    ylabel={\emph{Effective} GFLOPS ($ 2 \cdot \NIm{} \cdot \NJn{} \cdot \NPk{} / time $)},
    xmin=1,
    xmax=5.2,
    ymin=15,
    ymax=35,
    xtick={1, 2, 3, 4, 5},
    ytick={5, 10, 15, 20, 25, 28.36, 30, 35},
    grid=major,
    axis background/.style={fill=lightgray!20},
    mark size=0.8pt,
    cycle list name=jianyucolorlineratio,
    restrict y to domain=1:inf,
    legend style={
        at={(0.99,0.01)},
        anchor=south east,
        legend columns=1,
        font=\tiny,
        rounded corners=1pt,
        nodes={scale=0.75, transform shape},
        cells={anchor=west},
    },
    legend entries = {
        $ TBLIS            $\\
        $ One$-$level~ABC  $\\ 
        $ One$-$level~AB   $\\ 
        $ One$-$level~Naive$\\ 
        $ Two$-$level~ABC  $\\ 
        $ Two$-$level~AB   $\\ 
        $ Two$-$level~Naive$\\ 
    },
    ]
\addplot table[x=ratio,y=tblis,col sep=comma] {plotdata/ratio_1core.csv};
\addplot table[x=ratio,y=tensor_1_abc,col sep=comma] {plotdata/ratio_1core.csv};
\addplot table[x=ratio,y=tensor_1_ab,col sep=comma] {plotdata/ratio_1core.csv};
\addplot table[x=ratio,y=tensor_1_naive,col sep=comma] {plotdata/ratio_1core.csv};
\addplot table[x=ratio,y=tensor_2_abc,col sep=comma] {plotdata/ratio_1core.csv};
\addplot table[x=ratio,y=tensor_2_ab,col sep=comma] {plotdata/ratio_1core.csv};
\addplot table[x=ratio,y=tensor_2_naive,col sep=comma] {plotdata/ratio_1core.csv};
\end{axis}
\end{tikzpicture}
\quad
\begin{tikzpicture}[scale=0.9]
\begin{axis}[
    title={$ N_a = N_b = N_e = N_f $, $ N_i = N_j $, 10 core},
    xlabel={ $N_a \colon N_i$ ratio },
    ylabel={\emph{Effective} GFLOPS ($ 2 \cdot \NIm{} \cdot \NJn{} \cdot \NPk{} / time $)},
    xmin=1,
    xmax=5.2,
    ymin=120,
    ymax=280,
    xtick={1, 2, 3, 4, 5},
    ytick={50, 100, 120, 160, 200, 248, 280},
    grid=major,
    axis background/.style={fill=lightgray!20},
    mark size=0.8pt,
    cycle list name=jianyucolorlineratio,
    restrict y to domain=1:inf,
    legend style={
        at={(0.99,0.01)},
        anchor=south east,
        legend columns=1,
        font=\tiny,
        rounded corners=1pt,
        nodes={scale=0.75, transform shape},
        cells={anchor=west},
    },
    legend entries = {
        $ TBLIS            $\\
        $ One$-$level~ABC  $\\ 
        $ One$-$level~AB   $\\ 
        $ One$-$level~Naive$\\ 
        $ Two$-$level~ABC  $\\ 
        $ Two$-$level~AB   $\\ 
        $ Two$-$level~Naive$\\ 
    },
    ]
\addplot table[x=ratio,y=tblis,col sep=comma] {plotdata/ratio_10core.csv};
\addplot table[x=ratio,y=tensor_1_abc,col sep=comma] {plotdata/ratio_10core.csv};
\addplot table[x=ratio,y=tensor_1_ab,col sep=comma] {plotdata/ratio_10core.csv};
\addplot table[x=ratio,y=tensor_1_naive,col sep=comma] {plotdata/ratio_10core.csv};
\addplot table[x=ratio,y=tensor_2_abc,col sep=comma] {plotdata/ratio_10core.csv};
\addplot table[x=ratio,y=tensor_2_ab,col sep=comma] {plotdata/ratio_10core.csv};
\addplot table[x=ratio,y=tensor_2_naive,col sep=comma] {plotdata/ratio_10core.csv};
\end{axis}
\end{tikzpicture}
\caption{Performance for the contraction
$\T{Z}_{abij} := \T{W}_{abef} \cdot \T{T}_{efij}$ with varying $N_a \colon N_i$ ratio.
Left: performance on single core. Right: performance on one socket.
}
\label{fig:ratio}
\end{figure*}

\noindent
{\bf Synthetic tensor contractions.}
To evaluate the overall performance of various \strassen{} TC comparing against \tblis{} TC for different tensor problem sizes, shapes, and permutations,
we randomly generate
TC test cases with 2-D to 6-D randomly permuted tensors as operands, and test all these implementations for each synthetic test case.
We choose step size $256$ to sample uniformly $\{\NIm{}, \NJn{}, \NPk{}\}$ for various tensor bundle lengths:
\emph{square}: \SQUARE{}; \emph{rank-$\NPk{}$}: \RANKK{}; \emph{fixed-$\NPk{}$}: \FIXK{}.
For each bundle length
$\{\NIm{}, \NJn{}, \NPk{}\}$,
we randomly generate three $\{I_m, J_n, P_k\}$ 1-D, 2-D, or 3-D bundles, such that the product of each index length is close to 
$\{\NIm{}, \NJn{}, \NPk{}\}$. 
The order of $\{I_m, J_n, P_k\}$ is then randomly permuted.

The generated bundle lengths may not exactly match the original sampled bundle lengths.
When we plot the actual performance of these synthetic test cases,
we set
$ \NImCap{} = \NJnCap{} = \NPkCap{} =  (\NIm{} \cdot N_{J_n} \cdot N_{P_k} )^{1/3} $ for the \emph{square} bundle lengths;
$ \NPkCap{} = \NIm{} \cdot \NJn{} \cdot \NPk{} / (16000 \cdot 16000 ) $ for \emph{rank-$\NPk{}$} bundle lengths;
$ \NImCap{} = \NJnCap{} = ( \NIm{} \cdot \NJn{} \cdot \NPk{} / 1024 ) ^{1/2} $ for \emph{fixed-$\NPk{}$} bundle lengths.

For the \emph{square} and \emph{rank-$\NPk{}$} tensor shapes on one core, \tblis{} is rapidly outpaced by \ABCstrassen{}, with a crossover point of about $500 \approx 2\cdot k_C$.
\ABCstrassen\ is then shortly overtaken by \ABXstrassen\ and then by two-level \ABXstrassen. As predicted by the performance model, the \ABXstrassen\ implementation
is best for very large problem sizes due to repeated updates to $C$ in the \ABCstrassen\ algorithm. The \XXXstrassen\ implementations are never the best in these
experiments, although they may become more efficient than \ABXstrassen\ for extremely large, square problems. These trends are repeated in the ten-core experiments, although the crossover points are moved to larger tensor sizes.

For the \emph{fixed-$\NPk{}$} shapes, total performance
is lower for \ABXstrassen\ and \XXXstrassen\, with scalability for the algorithms being especially impacted by the relatively smaller $\NIm{}$ and $\NJn{}$ sizes.
For these shapes \ABCstrassen\ is always the fastest method above the crossover point with standard \tblis{}.

The actual performance data matches the predicted performance very well, with some variation due to the randomization of the tensor lengths and permutations.
Using these performance models, it may be possible to analytically decide on which algorithm to apply for a given tensor contraction to achieve the highest
performance, allowing an automated and seamless inclusion of \strassen\ into a TBLIS-like tensor framework.



\noindent
{\bf Real-world benchmark.}
In \figref{fig:benchmark},
we measure the performance of various implementations
for a subset of tensor contractions from the Tensor Contraction Benchmark \cite{TCB} on single core and one socket.
We present representative use cases where $\NPk{}$ is nearly equal to or larger than $2 k_C$ (512), for which
\strassen{} can show performance benefits, as illustrated in \secref{sec:model}.
The right three test cases represent various regularly-blocked tensor contractions from coupled cluster with single and double
excitations (CCSD)  \cite{shavitt_many-body_2009,helgaker_molecular_2013,scuseria_closed_shell_1987},
a workhorse quantum chemistry computational method. The fourth case from the
right illustrates the performance of \tblis{} and \strassen\ TC for a pure matrix case. Comparing
this case and the CCSD contractions highlights some of the performance issues that exist in the current
implementation of the packing and matrix-to-block scatter matrix copy kernels (see \secref{sec:packing} for
details). On one core, all \strassen\ implementations improve on \tblis{} for these right four cases, and in parallel
one-level \strassen\ implementations give a speedup as well, exceeding \ttt{} performance especially in the case of
\ABXstrassen. The gap between \tblis{} and \ttt{} for these contractions is due to \ttt{}'s use of Intel's MKL library, which is more
highly optimized than the BLIS/TBLIS framework.

The left two benchmarks are again quantum chemistry applications using 3-D tensors that arise in density-fitting (DF)
calculations \cite{DF1,DF2}. These contractions are also structurally equivalent to certain contractions from the coupled
cluster with perturbative triples, CCSD(T), method \cite{ccsd(t)}, where the \emph{occupied} (see \secref{sec:dist})
indices have been sliced. These cases show the improvement of \tblis{} over \ttt{} as noted in \cite{TC:Devin},
but do not show a speedup from \strassen\ except for one-level \ABCstrassen\ on one core. Our \strassen\ implementation
performs the submatrix multiplications sequentially, with only parallelization of each submatrix multiplication step. A more
comprehensive parallelization scheme, for example using task-based parallelism \cite{Benson15}, may show better performance. Additionally,
since the DF/CCSD(T) contractions are highly ``non-square'', an alternate fast matrix multiplication algorithm \cite{Benson15,FMM:IPDPS17} may
perform better.

%
%


\noindent
\textbf{Shape-dependence experiments.}
The performance of the ``particle-particle ladder'' contraction from CCSD, $\T{Z}_{abij} +\!\!= \T{W}_{abef} \cdot \T{T}_{efij}$
is reported for a range of tensor shapes in \figref{fig:ratio}. In these experiments, the length of the \emph{virtual} dimensions
$\{a,b,e,f\}$ is varied with respect to the length of the \emph{occupied} dimensions $\{i,j\}$ such that the total number of FLOPs
is roughly similarly to a $16000 \times 16000$ matrix multiplication, and the ratio $N_a \colon N_i$
is used as a proxy for tensor shape. A ratio of 1:1 would reflect an extremely poor quality of basis set for the overall
calculation, but is common when the calculation employs \emph{regular blocking}. The other end of the scale, with a ratio of
$\thicksim 5:1$, would then correspond to \emph{uneven blocking}. This type of blocking allows for better load balancing and
lower overhead when $N_a$ and $N_i$ are very unequal in the overall calculation.

The performance of \tblis{} and all of the
one-level \strassen\ algorithms show essentially no performance degradation across the entire range tested. The two-level
\strassen\ algorithms show some performance degradation at larger ratios, but still show improvement over \tblis{}. Eventually,
all \strassen\ algorithms will cross over and perform worse than \tblis{}, as evidenced by the left two contractions in \figref{fig:benchmark}
(these correspond to a ratio of about 22). However, the good performance of \strassen\ out to reasonably large ratios shows that
it could be beneficial in both regular blocking and uneven blocking scenarios.

\subsection{Distributed memory experiments}
\label{sec:dist}

We demonstrate how to use the \strassen{} TC implementations
to accelerate a distributed memory implementation of 4-D tensor contraction that exemplifies the two-particle ``ring'' terms from CCSD.
In our tests we set the  length of \emph{virtual} indices ($ abe $) to
$10\times$ that of \emph{occupied} indices ($ ijm $), which approximates the
use of a triple-$\zeta$ guality basis set. The problem sizes
tested here correspond to calculations on systems with 80, 112, 160, 192, and 224 electrons.
We use $\T{Z}_{abij} +\!\!=  \T{W}_{bmej} \T{T}_{aeim}$ as a demonstration example to show the performance benefit.

We implement a SUMMA-like\cite{SUMMA} algorithm for 4-D tensor contraction with MPI.
Initially the tensors $\T{W}$, $\T{T}$, and $\T{Z}$ are
distributed to a $ P \times P $ mesh of MPI processes using a 2D block distribution over the $a$, $b$, and $e$ dimensions, with the $i$, $j$,
and $m$ dimensions stored locally (i.e. not distributed).
After slicing $\T{W}$ and $\T{T}$ along the $e$ dimension,
the contraction is broken down into a sequence of
contractions of tensor slice pairs,
{
\begin{eqnarray*}
\T{Z} &+\!\!=&  \left(
\begin{array}{c | c | c }
\T{W}_{e;0} & \cdots & \T{W}_{e;K-1}
\end{array}
\right)
\left(
\begin{array}{c}
\T{T}_{e;0} \\ \hline
\vdots \\ \hline
\T{T}_{e;K-1}
\end{array}
\right)
\end{eqnarray*}%
}%
such that the $e$ index length for each tensor slice pairs $ \{\T{W}_{e;p}, \T{T}_{e;p} \}$
is $N'_e$.
For each tensor slice pairs, 
$ \T{W}_{e;p}$ is broadcast within rows of the mesh, and 
$ \T{T}_{e;p}$ is broadcast within columns of the mesh.
Then a local tensor contration for received tensor slice pairs is performed to update the local block.
Here \tblis{} TC and various \strassen{} TC are used as a drop-in replacement for this local tensor contraction.

\begin{figure}[htp!]
\center
\begin{tikzpicture}[scale=0.95]
\begin{axis}[
    title style={align=center},title={$\NIm{}(N_b \cdot N_j) = \NPk{}(N_e \cdot N_m) = \NJn{}(N_a \cdot N_i) \approx 16000 \cdot P$\\on $ P \times P $ MPI mesh\\1 MPI process per socket},
    xlabel={ $ P \times P $ },
    ylabel={\emph{Effective} GFLOPS ($ 2 \cdot  \NIm{} \cdot  \NJn{} \cdot \NPk{} / time $) / Socket},
    xmin=1,
    xmax=64,
    ymin=32,
    ymax=248,
    xtick={1,4,16,36,64},
    xticklabels={ {$ 1 $ }, $ 2 \times 2 $, $ 4 \times 4 $, $ 6 \times 6 $, $ 8 \times 8 $ },
    x tick label style={rotate=0,anchor=north,font=\tiny,align=center},
    ytick={0, 50, 100, 150, 200, 248},
    grid=major,
    axis background/.style={fill=lightgray!20},
    mark size=0.8pt,
    cycle list name=jianyucolordistline,
    restrict y to domain=1:inf,
    legend style={
        at={(0.99,0.01)},
        anchor=south east,
        legend columns=1,
        font=\tiny,
        rounded corners=1pt,
        nodes={scale=0.75, transform shape},
        cells={anchor=west},
    },
    legend entries = {
        $ TBLIS            $\\
        $ One$-$level~ABC  $\\ 
        $ One$-$level~AB   $\\ 
        $ One$-$level~Naive$\\ 
        $ Two$-$level~ABC  $\\ 
        $ Two$-$level~AB   $\\ 
        $ Two$-$level~Naive$\\ 
        $ CTF              $\\
    },
    ]
\addplot table[x=dim,y=tblis,col sep=comma] {plotdata/dist.csv};
\addplot table[x=dim,y=tensor_1_abc,col sep=comma] {plotdata/dist.csv};
\addplot table[x=dim,y=tensor_1_ab,col sep=comma] {plotdata/dist.csv};
\addplot table[x=dim,y=tensor_1_naive,col sep=comma] {plotdata/dist.csv};
\addplot table[x=dim,y=tensor_2_abc,col sep=comma] {plotdata/dist.csv};
\addplot table[x=dim,y=tensor_2_ab,col sep=comma] {plotdata/dist.csv};
\addplot table[x=dim,y=tensor_2_naive,col sep=comma] {plotdata/dist.csv};
\addplot table[x=dim,y=ctf,col sep=comma] {plotdata/dist.csv};
\end{axis}
\end{tikzpicture}
\caption{Weak scalability performance result of the various implementations for
a 4-D tensor contraction CCSD application
on distributed memory:
$\T{Z}_{abij} +\!\!=  \T{W}_{bmej} \T{T}_{aeim}$.
CTF shows the performance of 
the Cyclops Tensor Framework\cite{CTF2014} (linked with Intel MKL).
}
\label{fig:dist}
\end{figure}
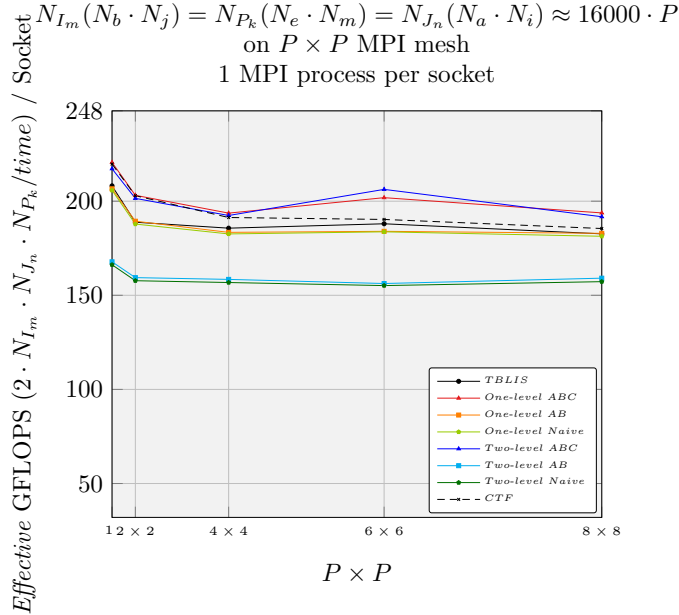

We perform the distributed memory experiment on the same machine as the single node experiment.
The dual-socket processor has ten cores on each socket. We run one MPI process for each socket,
and leverage all ten cores in a socket with thread parallelism for all implementations.
\figref{fig:dist} reports the weak scalability performance result on
up to 640 cores (32 nodes, 64 sockets).

In our experiments on $ P \times P $ mesh of sockets (MPI processes),
the lengths of virtual indices are set to equal $ N_a= N_b = N_e \approx 400 \sqrt{P} $
and the lengths of occupied indices are set to equal $ N_m= N_i = N_j \approx 40 \sqrt{P} $,
which make $\NIm{}=\NJn{}=\NPk{} \approx 16000 \cdot P$.
This guarantees the local memory buffer allocated to $\T{Z}$, $\T{W}$, $\T{T}$ is constant.
Our experiments verify that the above SUMMA-like algorithm is weakly scalable on this constant local memory setup,
regardless of which local TC implementation we use.
The local $e$ index length $N'_e$
is chosen close to  $N'_e = 1024/N_m$  such that the local TC computations are performed with
$\NPk{}= N'_e \cdot N_m \approx 4\cdot k_C$.
The tensor slice pairs in the local TC computations matches the shape
when \ABCstrassen{} achieves the best performance.
Therefore,
the one-level and two-level \ABCstrassen{} implementations outperform all other implementations.

We also tested the Cyclops Tensor Framework (CTF) \cite{CTF2014} which also uses a SUMMA or nested SUMMA algorithm but with possibly different block sizes and tensor distributions, as well as using the \ttdt\ algorithm for local tensor contractions.
We show it here as a reference for state-of-the-art performance.




%
%

\section{Conclusions}

We have presented what we believe to be the first paper to demonstrate how to leverage Strassen's algorithm for tensor contraction, and have shown practical performance speedup on single core, multicore, and distributed memory implementations.
Using a block scatter matrix layout enables us to partition the matrix view of the tensor, instead of the tensor itself, with automatic (implicit) tensor-to-matrix transformation,
and the flexibility to facilitate Strassen's 2D matrix partition to multi-dimensional tensor spaces.
Fusing the matrix summation that must be performed for \strassen{} and the transposition that must be conducted for tensor contraction
with the packing and micro-kernel operations inside high-performance implementation of GEMM
avoids extra workspace requirements,
and reduces the cost of additional memory movement.
We provided a performance model which can accurately predict the speedup of the resulting family of algorithms for different tensor shapes, sizes, and permutations.
We evaluated our families of implementations for various tensor sizes and shapes on synthetic and real-world datasets, both observing significant speedups
comparing to the baseline (\tblis{}) and naive implementations (\XXXstrassen{}),
particularly for smaller problem sizes ($\NIm{}, \NJn{}, \NPk{} \approx  2k_C,4k_C$), and irregular shape ($\NPk{}$ is much smaller comparing to $\NIm{}$, $\NJn{}$).
Together,
this work demonstrates Strassen's algorithm can be applied for tensor contraction with practical performance benefit.

There are several avenues for future work:

\begin{itemize}[leftmargin=*]
\item
Higher-level tensor decomposition algorithms \cite{kolda_tensor_2009}, such as Tucker decomposition, involve heavy use of
tensor contraction.
The impact of our performance improvements with Strassen's algorithm for those algorithms is an interesting question.
It may be possible to leverage our performance model to 
determine the best implementation for the tensor shape these algorithms require.

\item
So far, we target dense tensor contraction, which has numerous applications.
However, the structure of the tensor operands may be symmetric \cite{symmetrictensor}
or sparse \cite{sparsetensor}, which yields a number of new challenges, like more efficient storage or layout format.
How to explore those structure patterns and combine with Strassen's algorithm can be investigated.

\item
More levels of Strassen's algorithm may lose precision due to numerical instability issues.
It may be possible to combine with the techniques proposed in Extended and Mixed Precision BLAS
\cite{mixedBLAS} to get higher speedup and maintain precision.

\item 
A number of recent papers  explore practical implementations of Strassen-like fast matrix multiplications \cite{Benson15,FMM:IPDPS17}.
How to extend fast matrix multiplication with different partition block sizes for tensor contraction is an open question.





\end{itemize}



\section*{Additional information}

Additional information regarding BLIS and related projects can be found at 
\begin{center}
{\tt http://shpc.ices.utexas.edu}
\end{center}

\section*{Acknowledgments}

This work was sponsored in part by the National Science Foundation under grant number ACI-1550493, by Intel Corporation through an Intel Parallel Computing Center grant, and by a gift from Qualcomm. Access to the Maverick supercomputers administered by TACC is gratefully acknowledged.
DAM is an Arnold O. Beckman Postdoctoral Fellow.
We thank 
Martin Schatz for his help with distributed memory implementations, and
the rest of the SHPC team ({\tt http://shpc.ices.utexas.edu}) for their supports.

{\em Any opinions, findings, and conclusions or recommendations expressed in this material are those of the author(s) and do not necessarily reflect the views of the National Science Foundation. }

\bibliographystyle{IEEEtran}

\bibliography{biblio}


\end{document}